\PassOptionsToPackage{unicode}{hyperref}
\PassOptionsToPackage{hyphens}{url}
\documentclass[11pt]{article}
\usepackage{xcolor}
\usepackage[margin=1in]{geometry}
\usepackage{graphicx}
\usepackage{wrapfig}
\usepackage{float}
\usepackage{amsmath,amssymb}
\usepackage{newunicodechar}
\newunicodechar{κ}{\ensuremath{\kappa}}
\newunicodechar{λ}{\ensuremath{\lambda}}
\newunicodechar{ρ}{\ensuremath{\rho}}
\newunicodechar{≈}{\ensuremath{\approx}}
\newunicodechar{×}{\ensuremath{\times}}
\newunicodechar{−}{\ensuremath{-}}
\usepackage{iftex}
\ifPDFTeX
  \usepackage[T1]{fontenc}
  \usepackage[utf8]{inputenc}
  \usepackage{textcomp} 
\else 
  \usepackage{unicode-math} 
  \defaultfontfeatures{Scale=MatchLowercase}
  \defaultfontfeatures[\rmfamily]{Ligatures=TeX,Scale=1}
\fi
\usepackage{lmodern}
\ifPDFTeX\else
\fi
\IfFileExists{upquote.sty}{\usepackage{upquote}}{}
\IfFileExists{microtype.sty}{
  \usepackage[]{microtype}
  \UseMicrotypeSet[protrusion]{basicmath} 
}{}
\makeatletter
\@ifundefined{KOMAClassName}{
  \IfFileExists{parskip.sty}{%
    \usepackage{parskip}
  }{
    \setlength{\parindent}{0pt}
    \setlength{\parskip}{6pt plus 2pt minus 1pt}}
}{
  \KOMAoptions{parskip=half}}
\makeatother
\NewDocumentCommand\citeproctext{}{}
\NewDocumentCommand\citeproc{mm}{%
  \begingroup\def\citeproctext{#2}\cite{#1}\endgroup}
\makeatletter
 \let\@cite@ofmt\@firstofone
 \def\@biblabel#1{}
 \def\@cite#1#2{{#1\if@tempswa , #2\fi}}
\makeatother
\newlength{\cslhangindent}
\newlength{\csllabelwidth}
\newenvironment{CSLReferences}[2] 
 {\begin{list}{}{%
  \setlength{\itemindent}{0pt}
  \setlength{\leftmargin}{0pt}
  \setlength{\parsep}{0pt}
  \ifodd #1
   \setlength{\leftmargin}{\cslhangindent}
   \setlength{\itemindent}{-1\cslhangindent}
  \fi
  \setlength{\itemsep}{#2\baselineskip}}}
 {\end{list}}
\usepackage{calc}

\providecommand{\tightlist}{%
  \setlength{\itemsep}{0pt}\setlength{\parskip}{0pt}}
\usepackage{bookmark}
\IfFileExists{xurl.sty}{\usepackage{xurl}}{} 
\hypersetup{
  pdftitle={Incident-Data Robustness Analysis of the OWASP Top 10 for LLM Applications (2026): How a Community-Expert Ranking Holds Up Against a Large-Scale LLM Incident Corpus},
  hidelinks,
  pdfcreator={LaTeX via pandoc}}
\hypersetup{colorlinks=true,linkcolor=blue,urlcolor=blue,citecolor=blue}

\usepackage{fancyhdr}
\title{Incident-Data Robustness Analysis of the OWASP Top 10 for LLM
Applications (2026): How a Community-Expert Ranking Holds Up Against a
Large-Scale LLM Incident Corpus}
\author{Kyriakos ``Rock'' Lambros\\{\small OWASP GenAI Security Project
--- Top 10 for LLM Applications, Co-Lead} \and Steve
Wilson\\{\small OWASP GenAI Security Project --- Top 10 for LLM
Applications, Founder \& Lead}}
\date{2026}

\begin{document}
\maketitle
\begin{abstract}
The OWASP Top 10 for LLM Applications ranks the risks that a community
of security practitioners judges most important. We ask a narrower
question: checked against the record of real incidents, does that expert
ranking agree with the data? We assembled a large-scale corpus of
LLM-security incidents --- 7,714 snapshotted and 6,639 labeled against
the 20-entry taxonomy --- drawn from CVE, GHSA, OSV, and AIAAIC, and
derived an incident-based ranking with a Bayesian measurement-error
model that corrects each category's count for classifier precision and
recall. The 2026 list, published in August 2026 after this analysis was
run, blends the two signals at fixed weights, 0.75 on the expert vote
and 0.25 on the data, so the corpus corrects the consensus without
overturning it. The agreement between the two rankings is weak: Cohen's
κ ≈ 0.20, with a 90\% interval that crosses zero. The expert ranking is
nonetheless robust. A pre-registered bake-off of four frontier
classifiers returns no winner --- none beats the incidence floor's
balanced accuracy of 0.863 --- and a ground-truth check leaves the
floor's ordering (Spearman ρ = 0.918 against held-out truth) in place.
This is an exploratory analysis by two working-group members, not the
official OWASP release, and it does not supersede the official list or
process.
\end{abstract}

{
\setcounter{tocdepth}{2}
\tableofcontents
}
\section{How to Read This Report}\label{how-to-read-this-report}

This report checks the 2026 OWASP Top 10 for LLM Applications against a
record of real incidents, and it is written for a security professional
who has never taken a statistics course. The argument runs top to bottom
in plain language. Wherever a data-science term first appears, a short
sidebar box defines it:

\begin{quote}
\textbf{Sidebar --- example.} Sidebars look like this. Each one explains
a single term in plain language. Skip them when the term is already
familiar; the main argument reads without them.
\end{quote}

Every term defined in a sidebar reappears in the Glossary at the end, so
this doubles as a reference.

\textbf{Scope, in one line.} This is an incident-data analysis by two
working-group members. The OWASP GenAI LLM Top 10 2026 published in
August 2026 (\citeproc{ref-owasp2026llmtop10}{OWASP GenAI Security
Project 2026}). This report is the analysis that ran during that cycle,
not the published document, and it does not set or supersede the
official list. The full scope statement is at the end.

The structure:

\begin{itemize}
\tightlist
\item
  \textbf{Part I} explains what the OWASP LLM Top 10 is, where the
  incident corpus comes from, and how the expert vote and the data
  combine into one ranking.
\item
  \textbf{Part II} walks through what the incident data says, one step
  at a time: the corpus, the classifier, its measured accuracy, the
  Bayesian model, the incident-derived ranking, and where that ranking
  agrees and disagrees with the experts.
\item
  \textbf{Part III} stress-tests the ranking against four frontier
  classifiers and a ground-truth check.
\item
  \textbf{Glossary}, \textbf{Limitations}, \textbf{Data and Code
  Availability}, and \textbf{Scope} close the report.
\end{itemize}

\section{Part I --- The List and How It's
Made}\label{part-i-the-list-and-how-its-made}

\subsection{What the list is, and why check it against
incidents}\label{what-the-list-is-and-why-check-it-against-incidents}

The OWASP Top 10 for LLM Applications is a ranked list of the security
risks that matter most when software is built on large language models
(\citeproc{ref-owasp2025llmtop10}{OWASP GenAI Security Project 2024}).
It is an expert-consensus product: practitioners score each candidate
risk, and the scores aggregate into an ordered list, from Prompt
Injection at the top down to the tenth entry. Organizations use the list
to steer defensive effort: which risks get a control, which get a test
case, which get an audit line.

An expert vote is one way to rank risk. It captures informed judgment,
but judgment carries its own biases: what practitioners have read about
recently, and what last year's list anchored them to. We wanted to check
the vote against a second, independent signal, the pattern of incidents
that have actually happened. A risk that is common in the field should
leave a trail in public incident records. Comparing the two signals is
the whole of this report, and the result is a calibrated one: the
incident data agrees with the expert ranking only weakly (Cohen's κ ≈
0.20, with an interval that crosses zero), and the expert ranking is
robust: four frontier classifiers and a ground-truth check do not move
it.

Cataloging AI failures so the field can learn from them is established
practice. The AI Incident Database collects real-world AI harms to stop
the same mistakes recurring (\citeproc{ref-mcgregor2021}{McGregor
2021}), and the AI Risk Repository consolidates 777 risks drawn from 43
published taxonomies (\citeproc{ref-slattery2024}{Slattery et al.
2024}). Both describe the risk landscape. Neither scores an
expert-consensus ranking against the incident record, which is the gap
this report addresses.

\begin{quote}
\textbf{Sidebar --- incident corpus.} A corpus is a fixed, documented
collection of records assembled for analysis. Our incident corpus is a
snapshot of publicly reported LLM-security incidents, each one a short
text description of something that went wrong, pulled from public
databases on a fixed date so the analysis reproduces exactly.
\end{quote}

\subsection{Where the corpus comes
from}\label{where-the-corpus-comes-from}

The corpus holds 7,714 incidents as snapshotted, of which 6,639 carry a
usable label against the taxonomy. Those 6,639 divide into two strata:
6,297 security incidents and 342 ai-harm incidents. The security stratum
comes from three public vulnerability databases (CVE
(\citeproc{ref-cveprogram}{MITRE Corporation 2026}), GHSA
(\citeproc{ref-ghsa}{GitHub 2026}), and OSV (\citeproc{ref-osv}{Open
Source Security Foundation 2026})) and reads like engineering: advisory
and patch text for exploits and data-leakage bugs. The ai-harm stratum
comes from AIAAIC (\citeproc{ref-aiaaic}{AIAAIC 2026}), a database of
AI-related harms and controversies, and reads like journalism: news
summaries of algorithmic discrimination and deepfake misuse.

A second corpus corroborates the first at smaller scale. The OWASP
Agentic Security Initiative (\citeproc{ref-owaspagentic2025}{OWASP GenAI
Security Project 2025}) contributed 46 independently curated agentic-AI
incidents; classifying them against the same taxonomy reproduced 26\% of
the label assignments. That is a modest agreement rate on a small,
differently sourced set, and we report it as corroboration, not proof.
We describe the main corpus as large-scale (7,714 incidents is a large
sample for this field) and make no claim to being first.

\subsection{The 0.75 / 0.25 blend}\label{the-0.75-0.25-blend}

The 2026 list comes from neither signal alone. It comes from a weighted
blend of the two. The expert signal is a practitioner survey: about 29
respondents scored each candidate risk on importance, and the scores
aggregate to a median rank per entry. The data signal is the
incident-derived rank that Part II builds. The blend combines them at
fixed weights.

The two witnesses are combined in score space. Each risk's expert rank
and incident rate are put on a common scale, weighted three-quarters to
the vote and one quarter to the data, and blended per posterior draw. On
the data side the rates of a rolled-up child add to their parent, since
incidents accumulate. Three risks whose incident recall the corpus
cannot estimate --- Data and Model Poisoning, Vector and Embedding
Weaknesses, and Unbounded Consumption --- take their position from the
vote alone. The result is a distribution over positions, reported as
three tiers in the final section.

The weighting is deliberately lopsided, for a reason this report builds
toward. The list is a practitioner-consensus artifact, so the consensus
leads. The incident data is a useful but noisy corrective: it
under-detects unevenly, it is drawn from whatever reaches public
databases, and it agrees with the expert ranking only weakly. At a
quarter weight, the data is strong enough to move a risk a full tier
when the gap between the two signals is large, and too weak to overturn
the consensus on one imperfect corpus. That balance is the point of the
split: the data tugs, the consensus holds.

\subsection{What changed from 2025 to
2026}\label{what-changed-from-2025-to-2026}

The 2026 cycle works from a field of 20 candidate entries, not 10. Ten
are incumbents carried over from 2025 (LLM01 through LLM10)
(\citeproc{ref-owasp2025llmtop10}{OWASP GenAI Security Project 2024}).
Six are new candidates the working group is tracking but has held off
the published ten (the NEW-* entries). Four are narrower risks the group
folded into a broader incumbent rather than list separately (the ROLL-*
rollups); the incidents belonging to a rolled-up child still count, now
toward its parent.

\begin{figure}[htbp]
\centering
\includegraphics[width=0.85\textwidth]{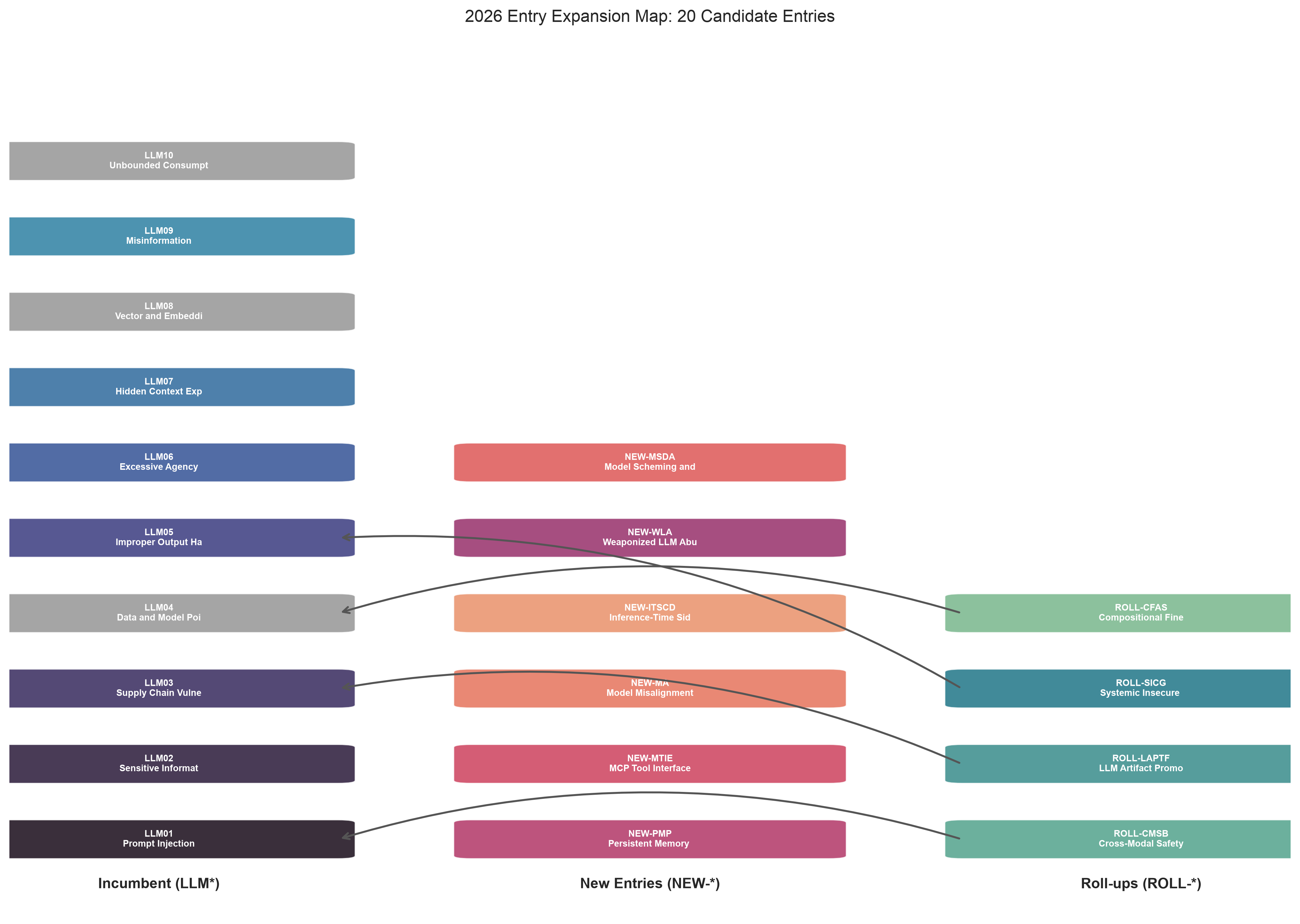}
\caption{The 2026 candidate field: ten incumbents, six new candidates, and four
rollups, each rollup arrowed to the incumbent it folds into.}
\end{figure}

Blending the two signals reorders the ten incumbents from their
published 2025 positions, and the reordering settles into the three
tiers this report uses throughout: a co-leading pair, a tied middle
band, and a wide tail. Excessive Agency is the one entry with a clear
stated move, entering the tied band. Every other incumbent's position
shifts inside its tier or holds; the model reports that movement as a
spread of plausible ranks per risk. The position chart in the blended
Top-10 section shows every incumbent's tier and its range of plausible
placements.

\section{Part II --- What the Incident Data
Says}\label{part-ii-what-the-incident-data-says}

\subsection{Act 1: The question}\label{act-1-the-question}

The 2025 OWASP Top 10 for LLM Applications came from expert consensus:
practitioner votes aggregated into a ranking, Prompt Injection at \#1,
Sensitive Information Disclosure at \#2, on down to \#10. This part
checks that consensus against a second signal, the incident record, one
step at a time.

We built a corpus of 6,639 labeled LLM-security incidents from public
databases, classified each against the 20-entry taxonomy, and derived a
data-driven ranking. The question is not whether the data proves the
experts right --- it cannot, and Part III shows the agreement is weak.
The question is what the data says on its own, and where it lines up
with the vote and where it does not.

Every chart and table below is computed live from the committed data;
re-run any cell to check it. The walkthrough covers how the
classification worked, how we measured its accuracy, and what a Bayesian
model does with noisy measurements. Each data-science term gets a
sidebar on first use.

The 20 taxonomy entries are listed below. The ``Incident Rank'' column
is blank for now; we fill it in Act 6, after the methodology.

\subsection{Act 2: The corpus}\label{act-2-the-corpus}

The corpus holds 6,639 labeled incidents in two strata that read very
differently. The security stratum (6,297 incidents from CVE, GHSA, and
OSV) is written like engineering: advisory text and patch notes for
prompt-injection exploits and data leakage through APIs. The ai-harm
stratum (342 incidents from AIAAIC) is written like journalism: news
summaries of algorithmic discrimination and deepfake misuse.

The split matters for what follows. The classifier reads the two strata
differently, and, as Act 4 shows, we could hand-verify its precision
only on the security stratum. Counts and corrections for ai-harm
categories therefore rest on weaker measurement than those for security
categories.

\begin{wrapfigure}{R}{0.45\textwidth}
\centering
\includegraphics[width=0.42\textwidth]{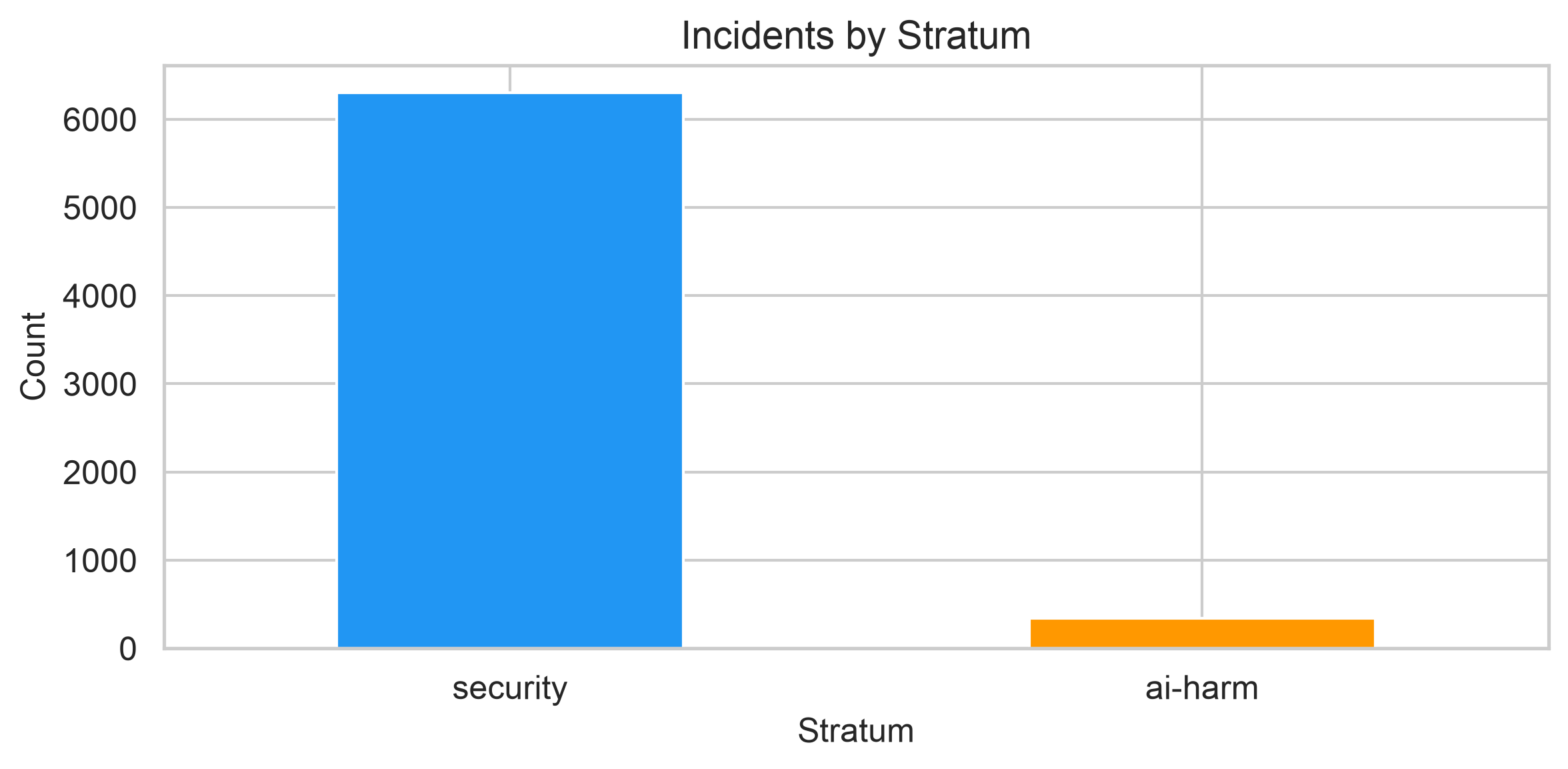}
\caption{Labeled incidents by stratum: the security sources (CVE, GHSA, OSV)
against the ai-harm source (AIAAIC).}
\end{wrapfigure}

\subsection{Act 3: Classifying 6,639
incidents}\label{act-3-classifying-6639-incidents}

Three large language models classified each incident independently: Qwen
235B (\citeproc{ref-qwen3}{{Yang et al.} 2025}), Llama 405B
(\citeproc{ref-llama3}{{Grattafiori et al.} 2024}), and DeepSeek V3
(\citeproc{ref-deepseekv3}{DeepSeek-AI 2024}). Each read the incident
text and assigned it to one of the 20 taxonomy entries, or marked it out
of scope when none fit. Using language models as annotators is now
common practice: on comparable labeling tasks they match or beat crowd
workers (\citeproc{ref-gilardi2023}{Gilardi et al. 2023}), and their
biases when used as judges are themselves documented
(\citeproc{ref-zheng2023}{Zheng et al. 2023}).

\begin{quote}
\textbf{Sidebar --- classifier.} A classifier is any procedure that
reads an input and assigns it to one of a fixed set of categories. Here
the input is an incident's text and the categories are the 20 taxonomy
entries plus ``out of scope.'' Our classifier is an ensemble of three
language models voting. It is a measuring instrument, not ground truth,
which is why Act 4 measures how often it is right.
\end{quote}

\begin{quote}
\textbf{Sidebar --- out-of-scope.} ``Out of scope'' is the category for
incidents that belong to none of the 20 taxonomy entries --- a real AI
harm that is not a vulnerability in a large language model, such as a
biased hiring tool or a surveillance drone. Marking an incident out of
scope is a correct classification, not a failure to classify.
\end{quote}

When all three models agreed on the same entry, we call it the agree
tier. When two agreed and one differed, the split tier. When all three
picked different entries, the disagree tier. The tier is a confidence
signal: agree-tier incidents have strong three-model consensus;
disagree-tier incidents sit in ambiguous territory where three
independent classifiers could not converge.

\begin{wrapfigure}{R}{0.43\textwidth}
\centering
\includegraphics[width=0.40\textwidth]{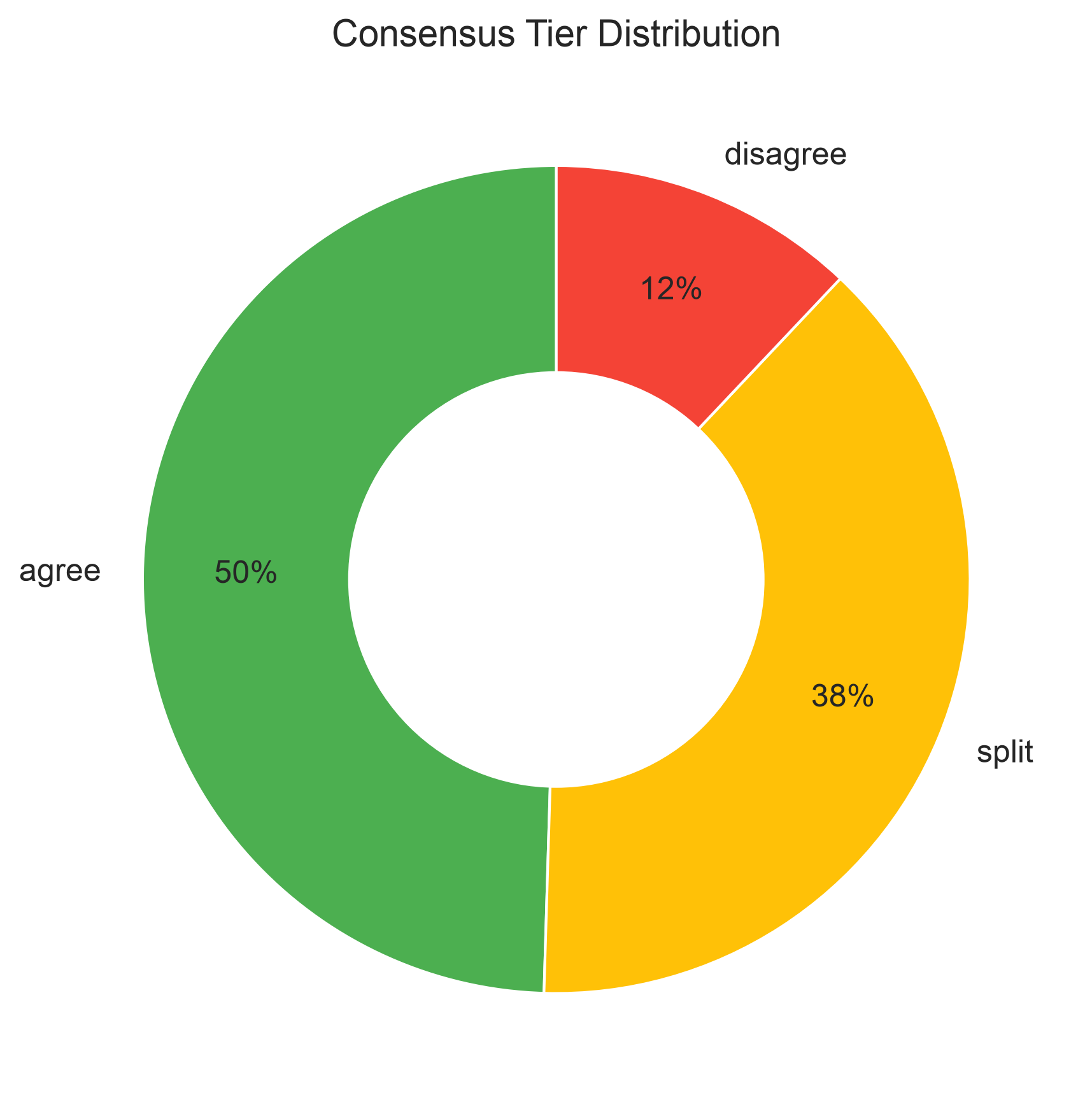}
\caption{Consensus tiers across the corpus: agree, split, and disagree.}
\end{wrapfigure}

\begin{figure}[htbp]
\centering
\includegraphics[width=0.72\textwidth]{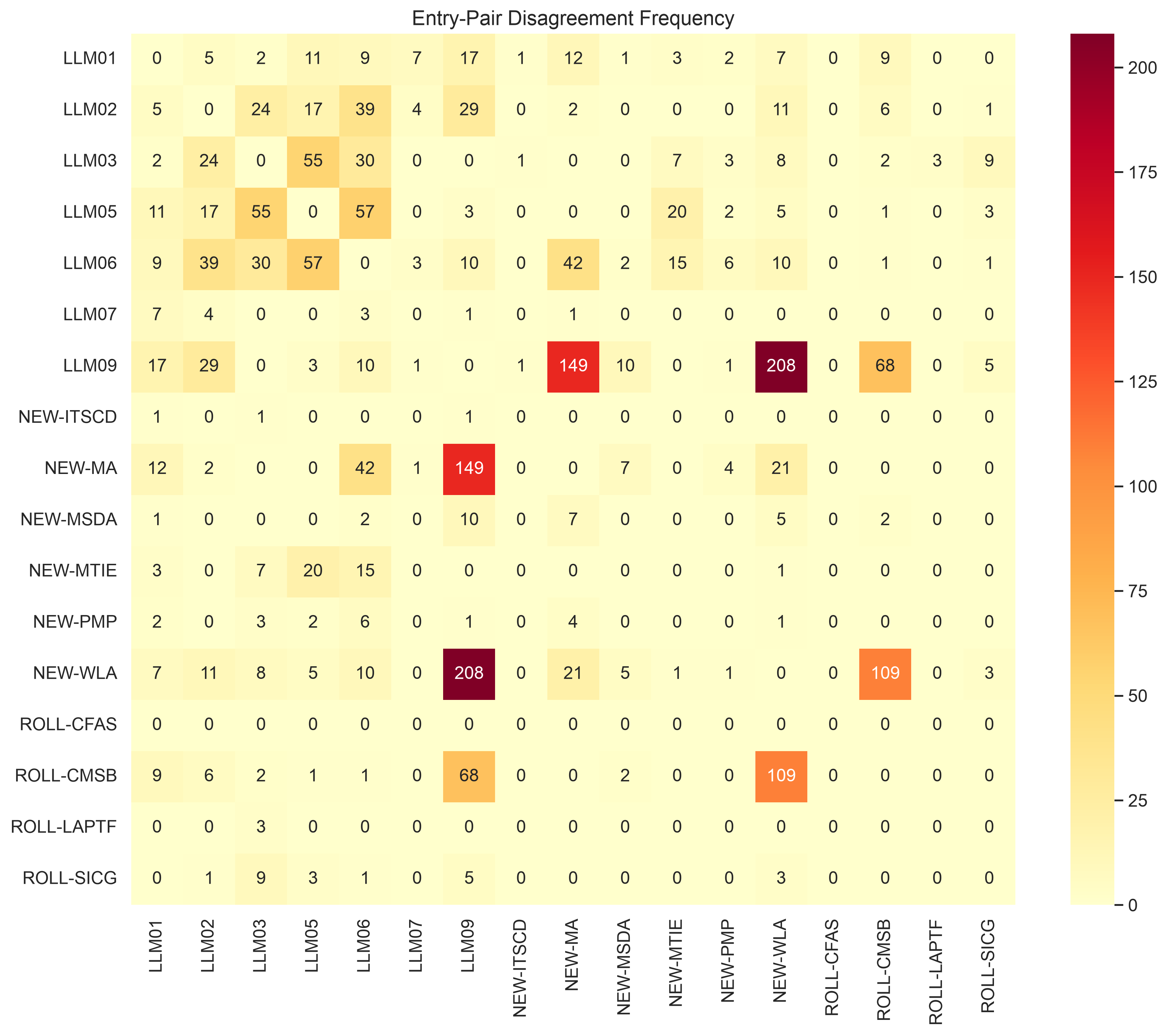}
\caption{Where the three classifiers disagree, by entry pair, across the
taxonomy.}
\end{figure}

\subsection{Act 4: How good is the
classifier?}\label{act-4-how-good-is-the-classifier}

The classifier is a measuring instrument, so we measured it. Two numbers
matter: how often its labels are correct, and how many true cases it
finds.

\begin{quote}
\textbf{Sidebar --- precision.} Of the incidents a classifier files
under a category, precision is the fraction that truly belong there. If
it labels 100 incidents ``LLM08'' and 13 of them really are LLM08, its
precision on LLM08 is 13\%. Low precision means most of what lands in a
category does not belong to it.
\end{quote}

\begin{quote}
\textbf{Sidebar --- recall.} Of the incidents that truly belong to a
category, recall is the fraction the classifier actually finds. A
classifier can have high precision and low recall (right when it fires,
but it rarely fires) or the reverse. Precision and recall answer
different questions and are measured separately.
\end{quote}

\begin{quote}
\textbf{Sidebar --- gold set.} A gold set is a batch of records labeled
carefully by a human to serve as the answer key. We use it two ways: to
measure the classifier's precision and recall, and, in Part III, as the
ground truth a ranking is scored against. Ours holds 1,200
human-adjudicated incidents.
\end{quote}

\begin{quote}
\textbf{Sidebar --- blind labeling.} Blind labeling means the human
records an independent judgment before seeing the machine's answer, so
the human is not anchored to it. Our reviewer labeled each incident from
its text alone, then revealed the three model votes, then made a final
decision. The blind step keeps the gold set from silently inheriting the
classifier's mistakes.
\end{quote}

We hand-verified 323 classifications as part of measuring precision, and
a reviewer adjudicated 1,200 incidents across all tiers to measure
recall. The precision posteriors combine those hand-verified checks with
the goldset adjudications. Precision varies sharply across entries, from
93\% (LLM01, LLM03) down to 13\% (LLM08). The variation is not random;
it tracks how cleanly each entry's definition separates it from its
neighbors. Four entries fall below the 50\% mark, each for a specific
reason.

\begin{itemize}
\tightlist
\item
  \textbf{LLM08 (Vector and Embedding Weaknesses), 13\%.} The lowest in
  the taxonomy: for every eight incidents labeled LLM08, about one
  belongs there. The category covers a narrow class of attacks on
  embedding spaces and vector stores, and the classifier confuses it
  with data-and-model-poisoning incidents (LLM04) and general
  data-integrity issues that are conceptually adjacent but taxonomically
  distinct.
\item
  \textbf{LLM07 (Hidden Context Exposure), 31\%.} The classifier
  struggles to separate exposing hidden system context (LLM07) from
  overriding it through injection (LLM01). Many real incidents involve
  both, because an attacker exposes the hidden context in order to craft
  a better injection. The boundary is clear in the taxonomy and blurred
  in practice.
\item
  \textbf{ROLL-CFAS (Compositional Fine-tuning Alignment Subversion),
  33\%.} Only one precision observation, so the posterior is dominated
  by its Beta(1,1) prior and the 90\% interval runs from 3\% to 78\%.
  The estimate says almost nothing; it reflects how little was measured,
  not a property of the category.
\item
  \textbf{ROLL-CMSB (Cross-Modal Safety Bypass), 44\%.} This entry sits
  on the confusion boundary examined in Act 9B. A deepfake that bypasses
  a content filter could read as cross-modal bypass (ROLL-CMSB),
  misinformation (LLM09), or weaponized abuse (NEW-WLA); the classifier
  picks one where a human might reasonably pick another.
\end{itemize}

Precision below 50\% means the classifier is wrong more often than right
for that entry. When you see an incident labeled LLM08, the odds are
about seven-to-one against it truly being a vector-or-embedding
weakness. This feeds directly into the Bayesian model in Act 5:
low-precision entries get large upward corrections, because much of
their observed count is misclassification noise, and wide uncertainty
ranges, because the correction itself is uncertain. A 13\% precision
estimate does not mean the entry is unimportant. It means the automated
measurement of that entry is unreliable, and the model's uncertainty
says so.

One coverage caveat. All 323 precision checks came from the security
stratum. The ai-harm stratum has no precision measurements, so the
Bayesian model uses a flat Beta(1,1) prior (prior mean 0.5) for ai-harm
precision. Error correction for ai-harm incidents rests on that
uninformative prior, not on direct measurement.

\begin{figure}[htbp]
\centering
\includegraphics[width=0.60\textwidth]{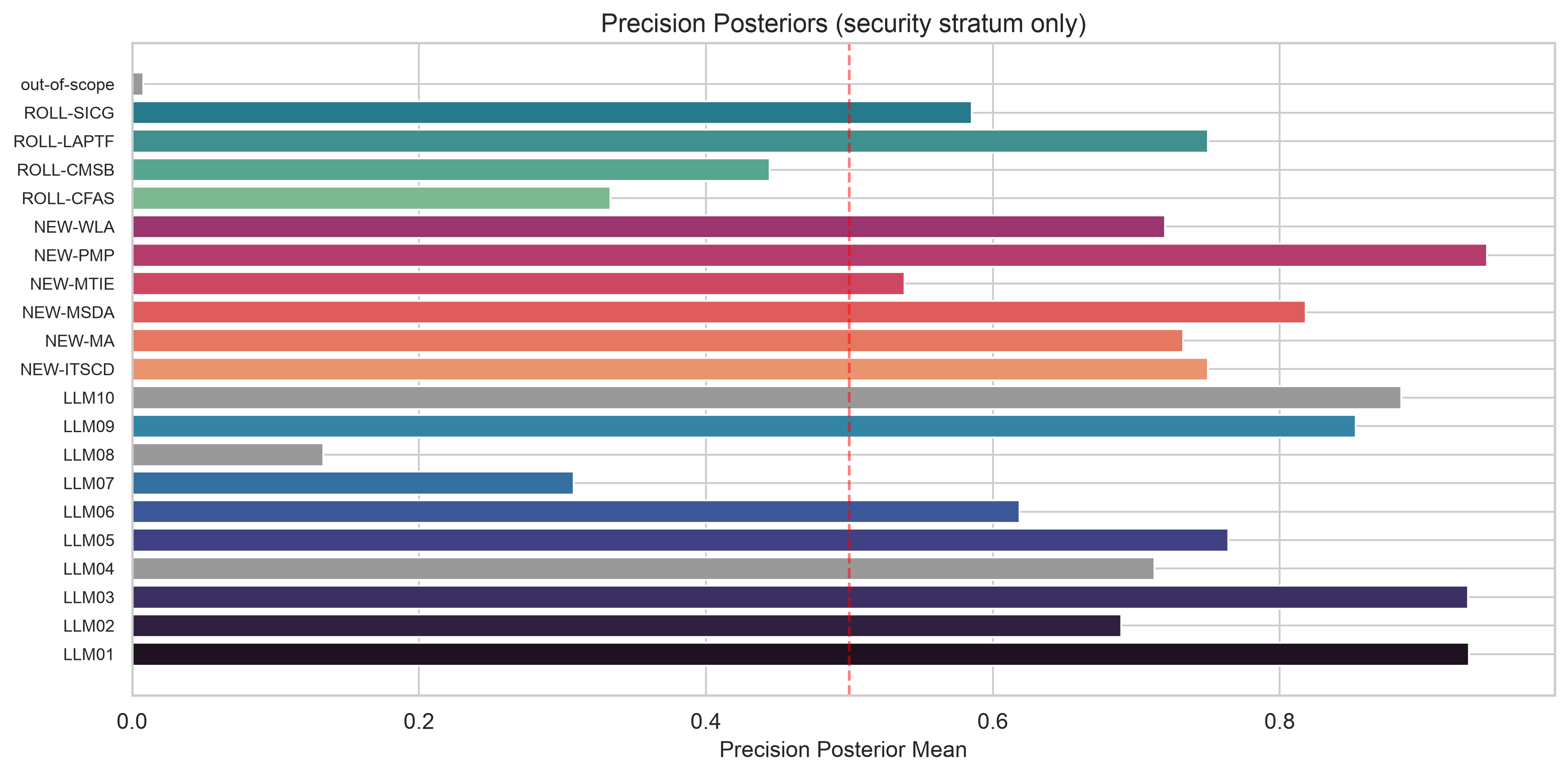}
\caption{Classifier precision by entry, with the 50\% line marked; precision
ranges from 93\% down to 13\%.}
\end{figure}

\begin{figure}[htbp]
\centering
\includegraphics[width=0.62\textwidth]{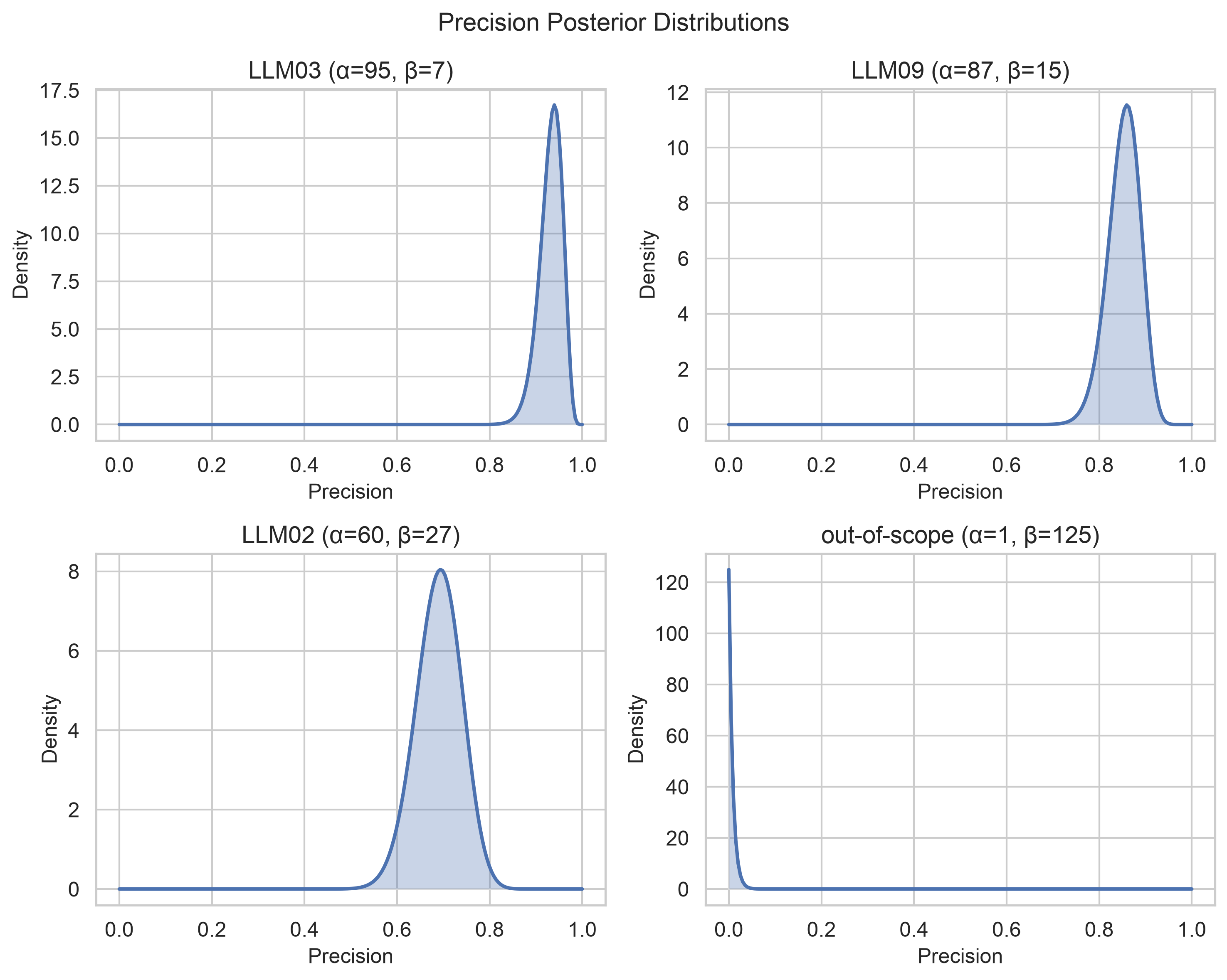}
\caption{Beta posteriors for per-entry precision; prior-dominated entries show
wide, flat curves.}
\end{figure}

\subsection{Act 5: From counts to rankings --- the Bayesian
model}\label{act-5-from-counts-to-rankings-the-bayesian-model}

Raw incident counts would mislead. An entry whose classifier runs at
30\% precision looks busy, but two-thirds of what lands there was
misclassified from somewhere else. We need a model that adjusts each
entry's observed count for its known classifier error and carries the
uncertainty of that adjustment through to the result. That is a
measurement-error model, the standard construction for inference when
the observed variable is a noisy proxy for the quantity of interest
(\citeproc{ref-carroll2006}{Carroll et al. 2006}).

The idea is a bathroom scale that reads two pounds heavy: you subtract
two pounds from every reading, and if the scale itself is uncertain (two
pounds off, give or take one) the corrected weight is uncertain too. The
model does this per entry, using the measured precision and recall from
Act 4.

\begin{quote}
\textbf{Sidebar --- latent incidence (λ).} Latent incidence, written λ
(lambda), is the quantity the model is really after: the true underlying
rate at which incidents of a category occur, as opposed to the raw count
the classifier reported. ``Latent'' means unobserved: we never see λ
directly, and infer it from the noisy counts after correcting for
precision and recall.
\end{quote}

\begin{quote}
\textbf{Sidebar --- prior and posterior.} A prior is what the model
assumes about a quantity before looking at this data; the posterior is
the updated belief after combining the prior with the data. The
posterior is not a single number but a distribution, a range of
plausible values with more weight on the likelier ones. A wide posterior
means the data left the answer uncertain
(\citeproc{ref-gelman2013}{Gelman et al. 2013}).
\end{quote}

\begin{quote}
\textbf{Sidebar --- negative-binomial measurement-error model.} Two
ideas in one name. ``Measurement-error model'' means the model treats
the classifier's counts as noisy measurements of the true rate and
corrects for the noise. ``Negative-binomial'' is the count distribution
it uses; unlike the simpler Poisson, it lets the spread of counts exceed
their average, which real incident counts do. Together: a count model
that expects over-dispersed data and corrects for classifier error.
\end{quote}

For entries above 50\% precision, the correction is a moderate downward
nudge: some observed incidents were misclassified in, so the true count
is a little lower. For entries below 50\% precision, the correction is
larger than the reading itself: if only 13\% of LLM08 labels are real,
the model must recover the true rate from a signal that is mostly noise.
Two things follow. The corrected estimate can sit far from the raw
count, and the uncertainty around it is wide, because small changes in
the precision estimate swing the corrected rate a lot. That is why some
entries in Act 6 span ten or more rank positions: the width is the model
honestly reporting how little the data pins that entry down.

\begin{quote}
\textbf{Sidebar --- MCMC.} Markov chain Monte Carlo is a way to explore
a probability distribution too complex to write down in closed form. It
takes a guided random walk through the space of possible answers,
spending more time where the answer is more plausible; the collected
steps approximate the posterior. We drew 16,000 samples this way (four
chains of 4,000, after 2,000 warm-up steps each), using the No-U-Turn
sampler (\citeproc{ref-hoffman2014}{Hoffman and Gelman 2014}) as
implemented in NumPyro (\citeproc{ref-phan2019}{Phan et al. 2019}). The
convergence checks beside the model output use the rank-normalized R-hat
and effective-sample-size diagnostics
(\citeproc{ref-vehtari2021}{Vehtari et al. 2021}).
\end{quote}

Three entries (LLM04, LLM08, LLM10) are frame-blind: their incidents
come almost entirely from one stratum, so the model cannot cross-check
their rates across strata. They stay in the analysis but are flagged,
and their rank estimates carry structural uncertainty beyond what the
intervals show.

Two measurement gaps widen the intervals further. For 16 of 20 entries
the ai-harm recall is not measured directly, so the model uses a
conservative prior (roughly 1\% recall, a Beta(1, 101)) and corrects
upward accordingly. And ai-harm precision is unmeasured, so the model
uses a flat Beta(1,1) prior there. Both choices are honest about missing
data, and both add width to the posteriors in Act 6.

\begin{figure}[htbp]
\centering
\includegraphics[width=0.70\textwidth]{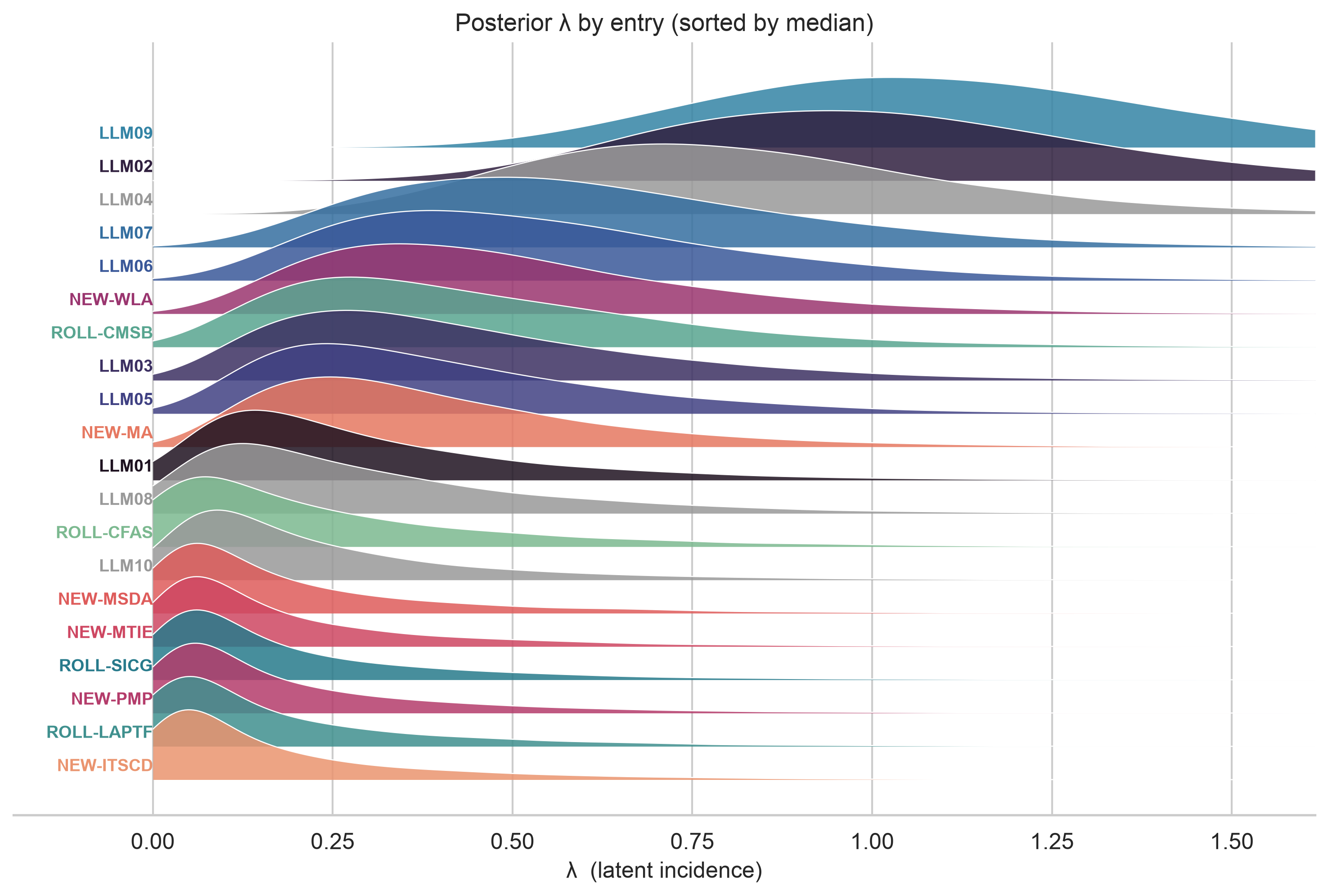}
\caption{Posterior distributions of latent incidence (λ) by entry.}
\end{figure}

\subsection{Act 6: The incident-derived
ranking}\label{act-6-the-incident-derived-ranking}

This ranking is what the incident data suggests after correcting for
classifier error. It is one signal of two, not the final word --- the
blend in Part I gives it a quarter weight, and Part III shows why that
restraint is warranted.

For each entry the model gives a posterior over its true incidence, and
we rank entries by their median. Each rank comes with a 90\% credible
interval.

\begin{quote}
\textbf{Sidebar --- credible interval.} A credible interval is the
Bayesian answer to ``how sure are we?'' A 90\% credible interval is the
range holding 90\% of the posterior's plausible values, so there is a
90\% chance the true value sits inside it, given the model and the data.
Wide intervals mean low certainty. When an entry's rank interval spans
ten positions, the data barely constrains where it belongs.
\end{quote}

How to read the chart: each row is an entry, the diamond marks its
median rank, and the bar spans the 90\% credible interval on that rank.
Tight intervals (LLM02 spans roughly 1--6) mean the data constrains the
position well. Wide intervals (spanning 6--20) mean the data is
compatible with many positions, which happens when precision is low,
observations are few, or recall is unmeasured. Grey entries (LLM04,
LLM08, LLM10) are frame-blind and carry extra structural uncertainty.

\begin{figure}[htbp]
\centering
\includegraphics[width=0.68\textwidth]{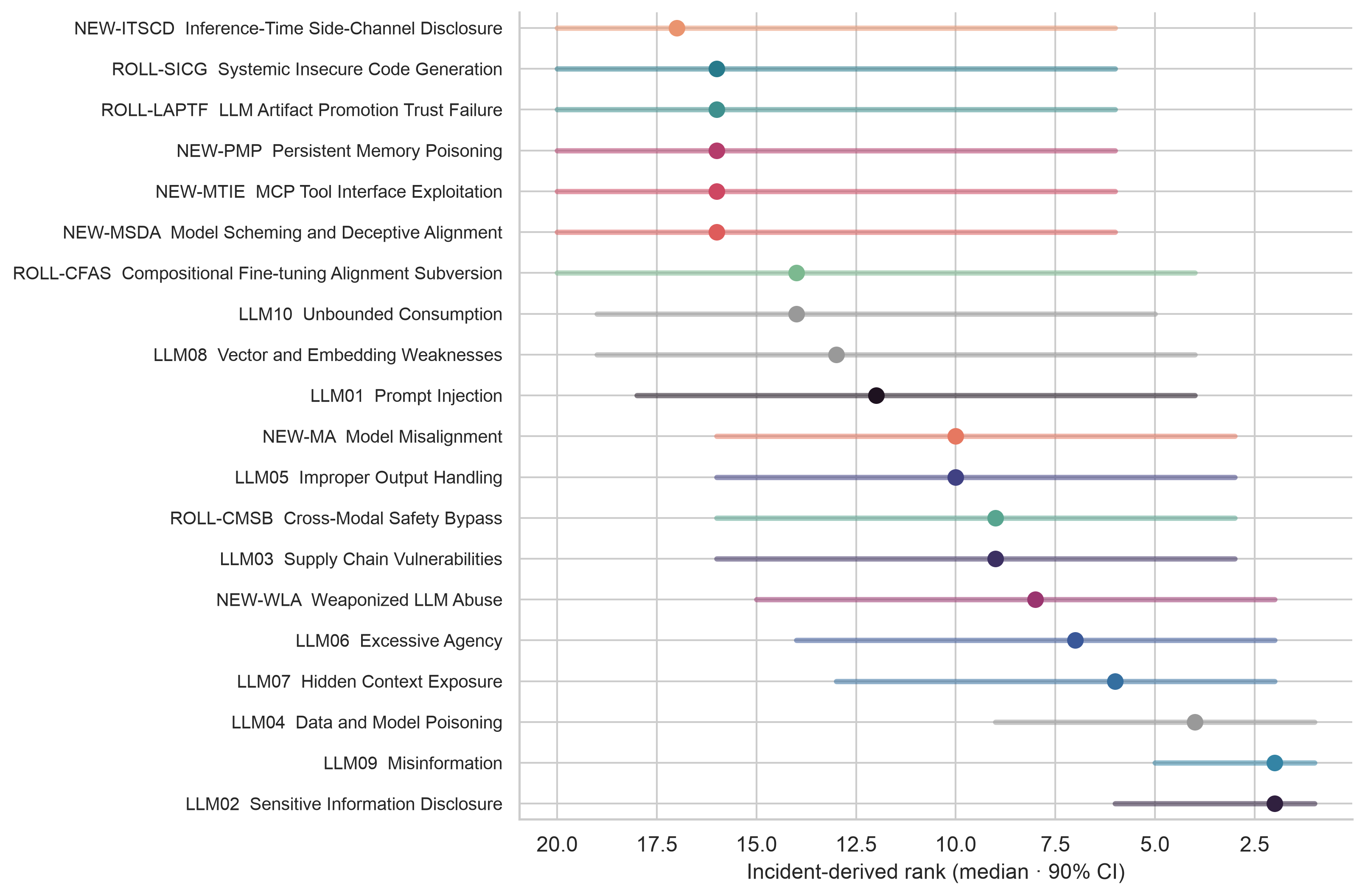}
\caption{Incident-derived rank by entry: median rank (diamond) and 90\% credible
interval (bar).}
\end{figure}

\subsection{Act 7: Do the experts and the incidents
agree?}\label{act-7-do-the-experts-and-the-incidents-agree}

\begin{quote}
\textbf{Sidebar --- Cohen's κ.} Cohen's kappa (κ) measures agreement
between two labelings after subtracting the agreement expected from
chance alone (\citeproc{ref-cohen1960}{Cohen 1960}). κ = 1 is perfect
agreement, κ = 0 is no better than chance, and negative κ is systematic
disagreement. The weighted version used here penalizes near-misses less
than far-misses (\citeproc{ref-cohen1968}{Cohen 1968}). κ comes with an
uncertainty interval, and that is the part that matters most: when the
interval crosses zero, the data cannot rule out chance-level agreement,
so the agreement is weak whatever the point estimate says.
\end{quote}

The agreement is weak. Comparing the expert ranking with the incident
ranking gives Cohen's weighted κ = 0.20, with a 90\% interval of −0.16
to 0.57. The interval crosses zero. We cannot exclude chance-level
agreement, and the point estimate of 0.20 sits only in the ``slight''
band of the conventional interpretation scale
(\citeproc{ref-landis1977}{Landis and Koch 1977}). This is the honest
headline of the whole analysis: the incident data agrees with the expert
ranking only weakly.

Two things make the interval wide. Only 17 of the 20 entries are
measurable (three are frame-blind), and agreement statistics need larger
samples to tighten. And the posterior rank distributions are themselves
wide, most spanning ten or more positions, which propagates into the
agreement estimate.

Five entries disagree the most. Across the joint posterior, the two
signals place each of them in different thirds of the ranking more than
83\% of the time:

\begin{itemize}
\tightlist
\item
  \textbf{LLM01 Prompt Injection:} experts \#1 (interval 1--2),
  incidents \#12 (4--18).
\item
  \textbf{LLM09 Misinformation:} incidents \#2 (1--5), experts \#13
  (9--16).
\item
  \textbf{NEW-MTIE MCP Tool Interface Exploitation:} experts \#7 (5--9),
  incidents \#16 (6--20).
\item
  \textbf{NEW-PMP Persistent Memory Poisoning:} experts \#4 (2--7),
  incidents \#16 (6--20).
\item
  \textbf{NEW-WLA Weaponized LLM Abuse:} incidents \#8 (3--15), experts
  \#17 (13--20).
\end{itemize}

\begin{figure}[htbp]
\centering
\includegraphics[width=0.72\textwidth]{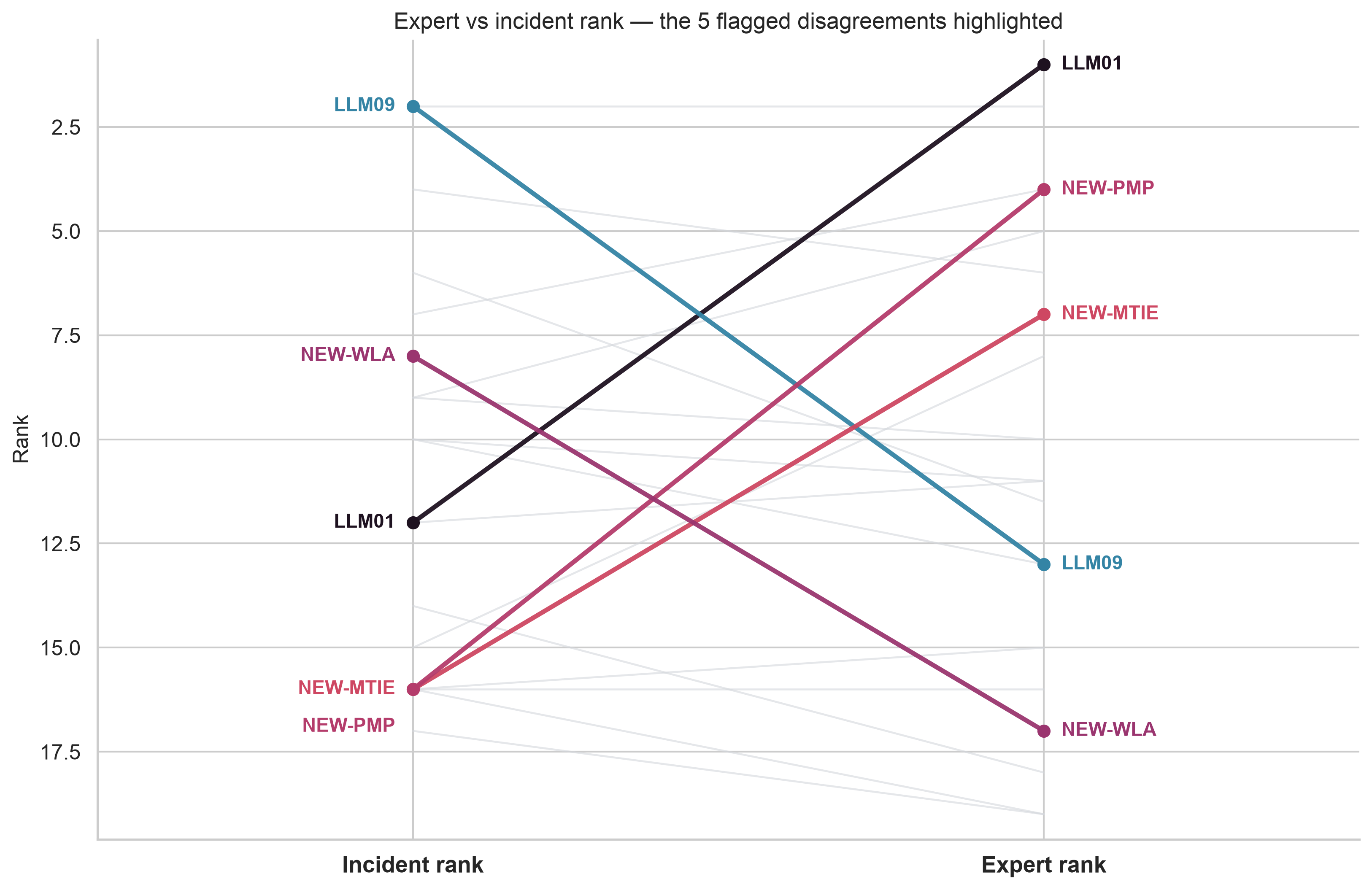}
\caption{Expert rank against incident rank, entry by entry.}
\end{figure}

\begin{figure}[htbp]
\centering
\includegraphics[width=0.72\textwidth]{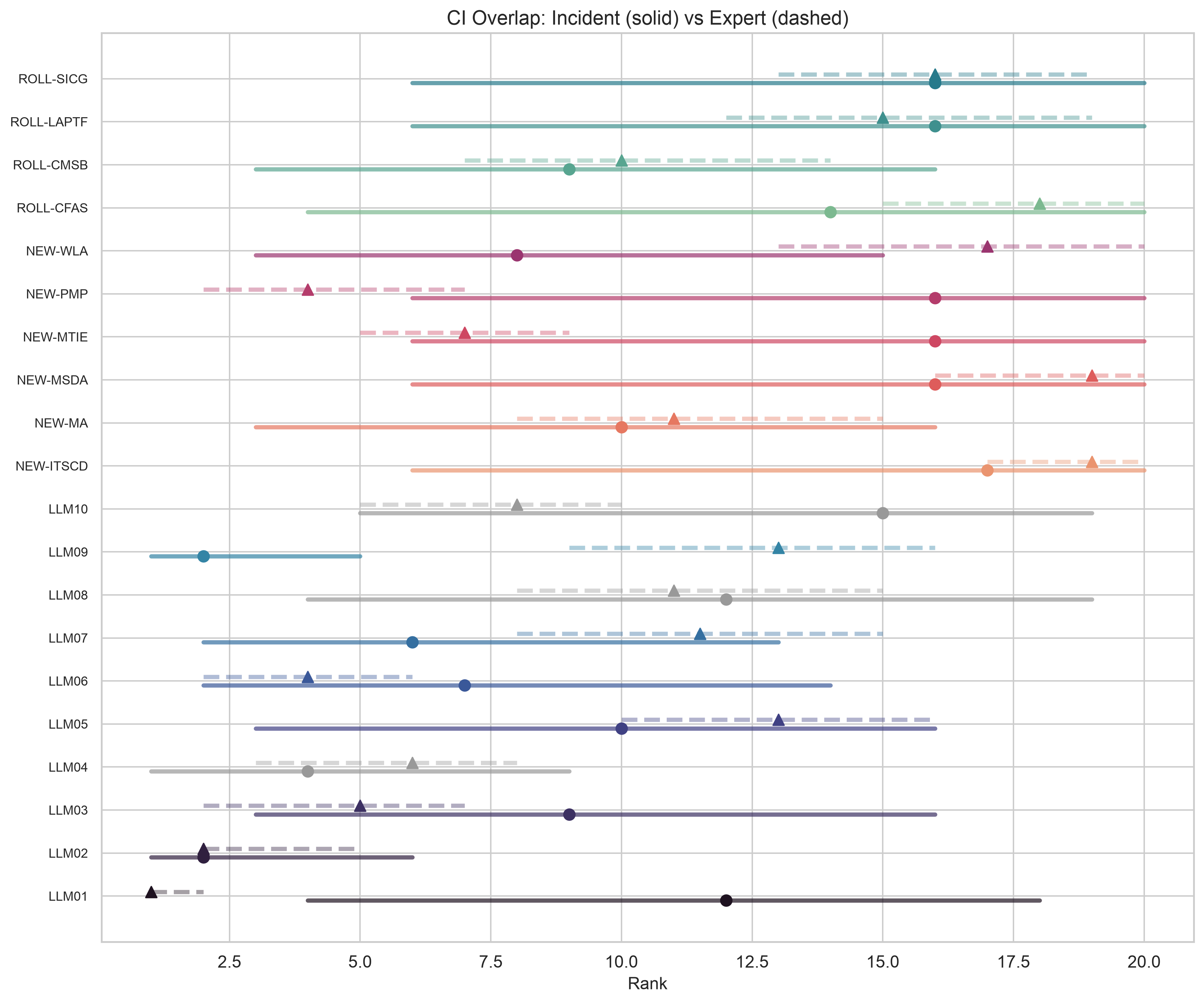}
\caption{Overlap of the expert and incident rank intervals, per entry.}
\end{figure}

\subsection{Act 8: Where the experts and the incidents
disagree}\label{act-8-where-the-experts-and-the-incidents-disagree}

Each of the five mismatches has a plausible cause. Reading them is more
useful than averaging them away.

\textbf{LLM01 (Prompt Injection): expert \#1, incident \#12.} Prompt
injection is the best-understood LLM attack
(\citeproc{ref-perez2022}{Perez and Ribeiro 2022};
\citeproc{ref-greshake2023}{Greshake et al. 2023}), and deployed systems
defend against it actively. Successful, publicly reported exploits are
correspondingly rarer, so the incident record sees fewer of them.
Experts rank it first because the attack surface stays enormous even
when the defenses mostly hold; the data sees the successes that got
through.

\textbf{LLM09 (Misinformation): incident \#2, expert \#13.} The corpus
carries a large volume of deepfake and AI-generated disinformation from
the AIAAIC harm database. Experts may rank it lower because the category
overlaps others (NEW-WLA, ROLL-CMSB) and because many of these incidents
describe harm produced by an AI rather than a vulnerability inside an
LLM. Act 9B examines that overlap.

Misinformation is the widest disagreement between the two witnesses. The
incident record ranks it near the top; the expert vote ranks it near the
bottom. The engine's concordance flag puts the probability that the two
signals disagree at 99 percent. That number quantifies disagreement
between the two signals. It does not quantify how underrated the risk
is, and the incident signal behind it rests on the ai-harm stratum,
whose precision the corpus does not measure directly. Read it as the
entry the incident record most disputes, and the one a better-measured
corpus is most likely to move.

\textbf{NEW-PMP (Persistent Memory Poisoning) and NEW-MTIE (MCP Tool
Interface Exploitation): expert top-5, almost no incidents.} These are
emerging threats whose public incident record has not caught up. If the
list exists to warn practitioners, expert signal should outweigh
incident counts for threats that are new by definition.

\textbf{NEW-WLA (Weaponized LLM Abuse): 863 incidents, expert \#17.} The
large count follows from a broad definition that absorbs AI-generated
disinformation, synthetic-media abuse, and deepfake harm. Experts rank
it low because most of these describe harm from an AI system rather than
an exploitable weakness in an LLM.

\begin{wrapfigure}{R}{0.49\textwidth}
\centering
\includegraphics[width=0.46\textwidth]{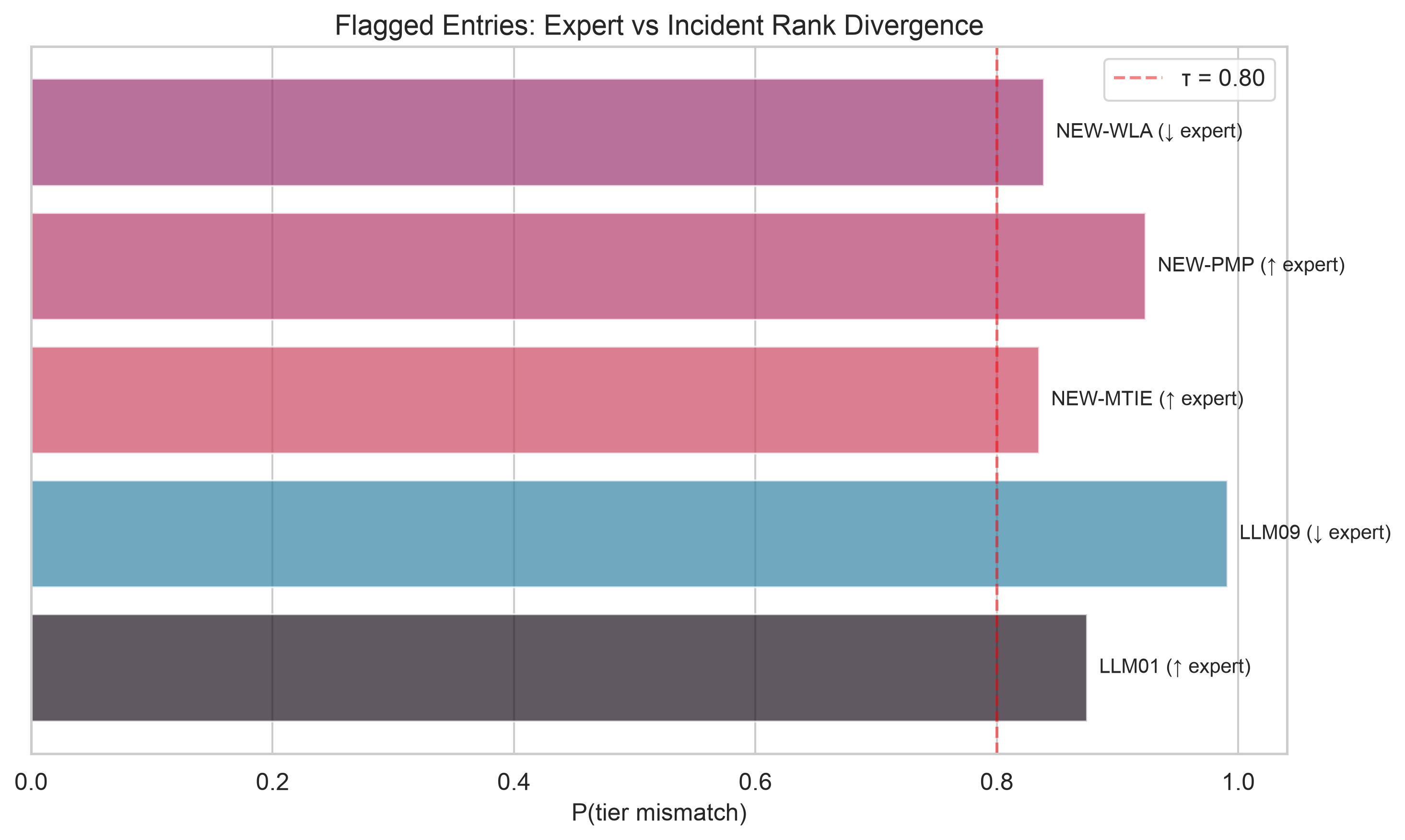}
\caption{The five largest expert-versus-incident tier mismatches.}
\end{wrapfigure}

\begin{wrapfigure}{L}{0.45\textwidth}
\centering
\includegraphics[width=0.42\textwidth]{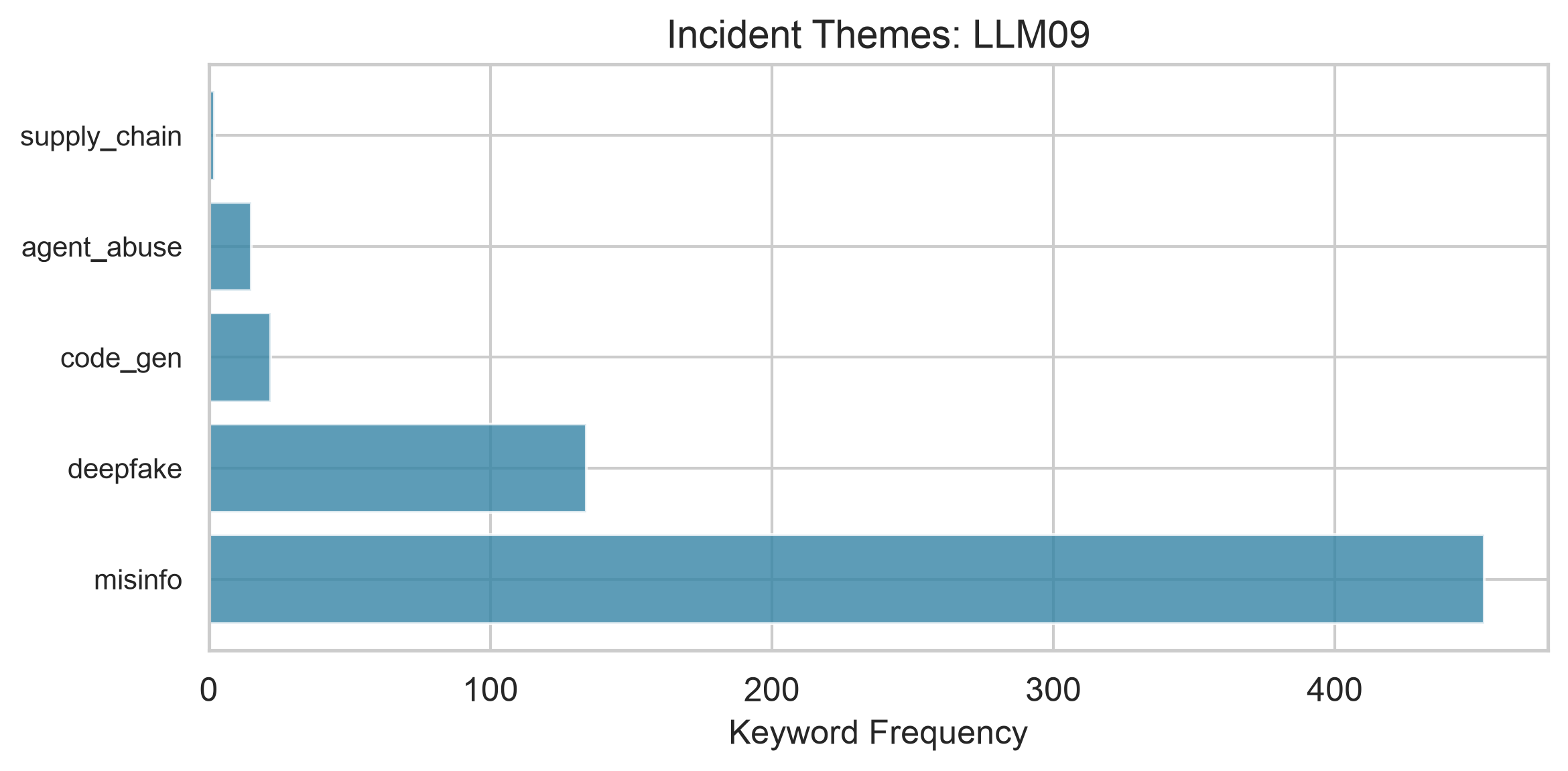}
\caption{Frequent keywords in LLM09 (Misinformation) incidents.}
\end{wrapfigure}

\begin{wrapfigure}{R}{0.45\textwidth}
\centering
\includegraphics[width=0.42\textwidth]{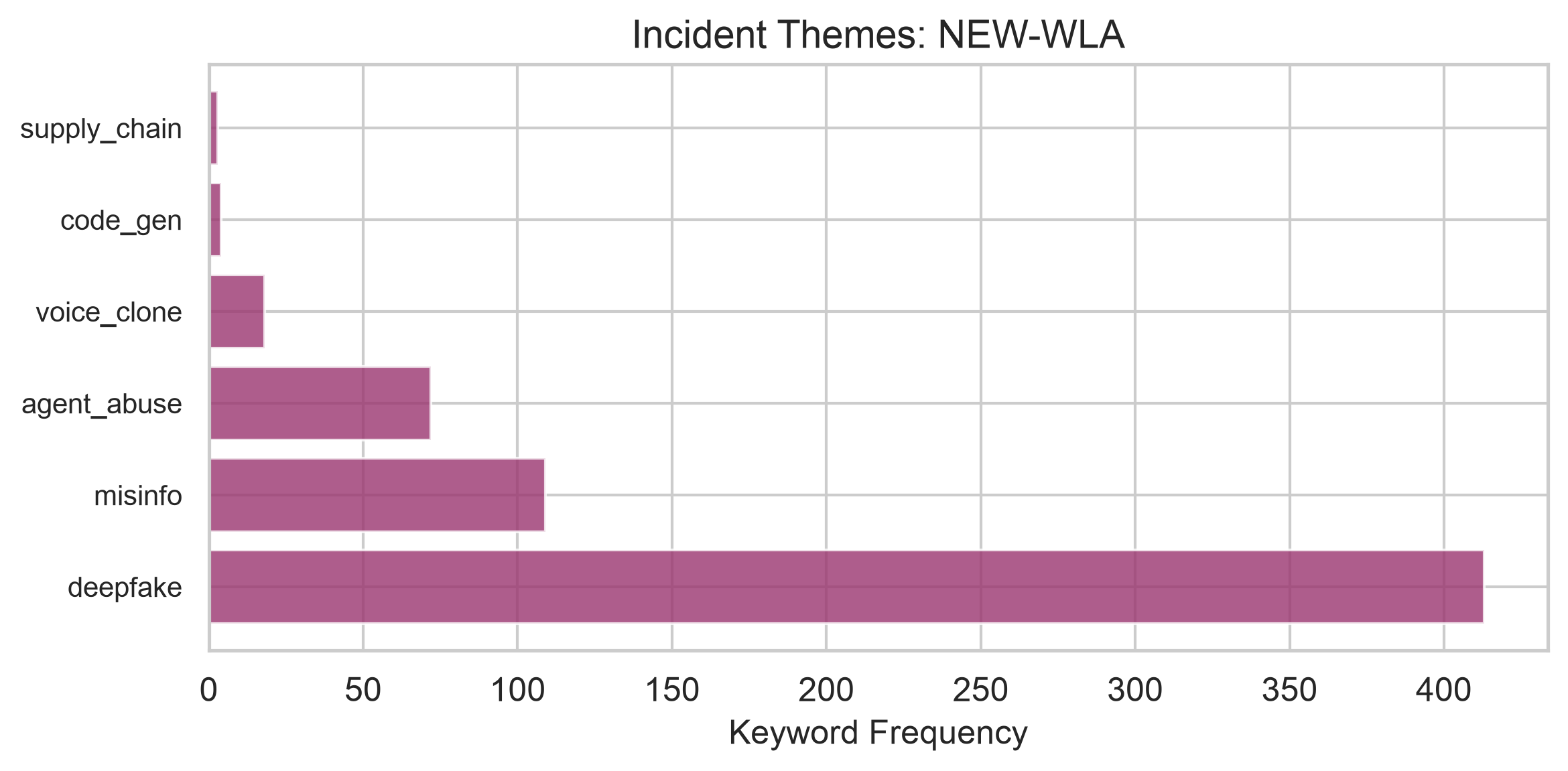}
\caption{Frequent keywords in NEW-WLA (Weaponized LLM Abuse) incidents.}
\end{wrapfigure}

\subsection{Act 9: What the data cannot
see}\label{act-9-what-the-data-cannot-see}

\subsubsection{9A: AI harm without an LLM
vulnerability}\label{a-ai-harm-without-an-llm-vulnerability}

More than a third of the labeled corpus --- 2,394 incidents --- landed
out of scope, where the models' consensus placed them. These are real AI
harms: facial recognition that misidentifies people, hiring tools that
discriminate, recommendation engines that radicalize. None describes a
vulnerability inside a large language model. They are harms from AI
systems, not vulnerabilities of LLMs.

The gap is a property of the sampling frame, not a failure of the
taxonomy. The corpus was built by crawling CVE, GHSA, and OSV with
AI-related keywords, which pull in anything mentioning ``AI'' or
``machine learning'' whether or not an LLM is involved, and AIAAIC
covers all AI harms by design. Where an incident sits outside the
taxonomy --- because it involves non-LLM AI, or describes a societal
effect rather than a technical weakness --- the incident signal is
silent. The out-of-scope pile marks the boundary of what this method can
measure.

\begin{figure}[htbp]
\centering
\includegraphics[width=0.85\textwidth]{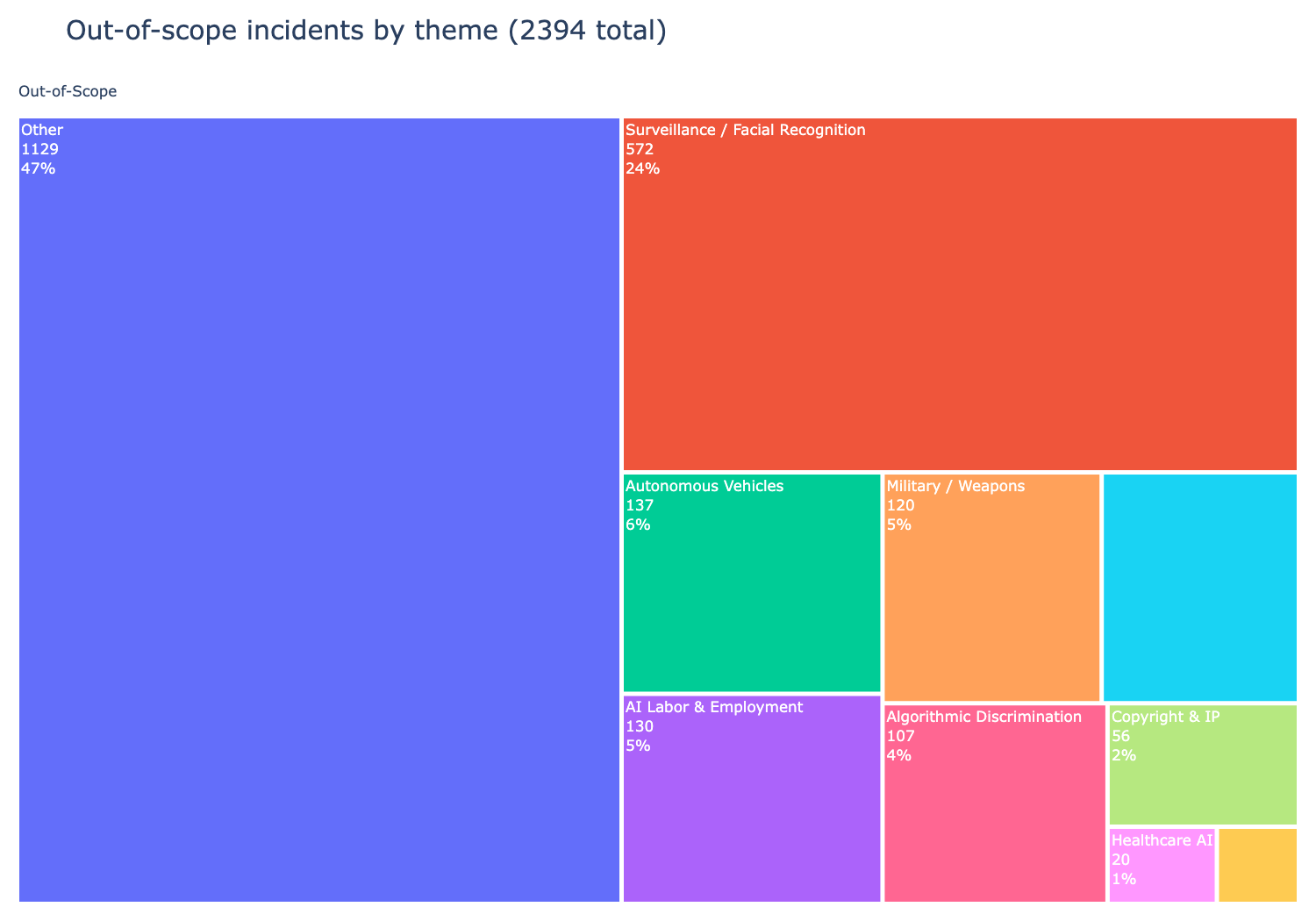}
\caption{Themes among the out-of-scope incidents.}
\end{figure}

\subsubsection{9B: The LLM09 / NEW-WLA / ROLL-CMSB confusion
boundary}\label{b-the-llm09-new-wla-roll-cmsb-confusion-boundary}

Some categories overlap enough that neither classifiers nor humans can
reliably tell them apart. That is a confusion boundary, and it is a
property of the categories, not a broken classifier. Three entries share
one:

\begin{itemize}
\tightlist
\item
  \textbf{LLM09 (Misinformation):} the output is false or misleading.
\item
  \textbf{NEW-WLA (Weaponized LLM Abuse):} an adversary uses AI as a
  weapon.
\item
  \textbf{ROLL-CMSB (Cross-Modal Safety Bypass):} the attack runs
  through image, video, or audio.
\end{itemize}

A deepfake video spreading political disinformation is all three at
once: misleading content, created as a weapon, through a visual
modality. The overlap is genuine ambiguity in the taxonomy, not a
labeling mistake. So when the incident data ranks LLM09 second, part of
that signal comes from incidents that could as easily have been filed
under NEW-WLA or ROLL-CMSB. The boundary inflates whichever entry the
classifier happens to prefer and deflates the others. The Bayesian model
corrects for measured precision, but it cannot correct for ambiguity the
human reviewers themselves found hard to resolve.

\begin{figure}[htbp]
\centering
\includegraphics[width=0.90\textwidth]{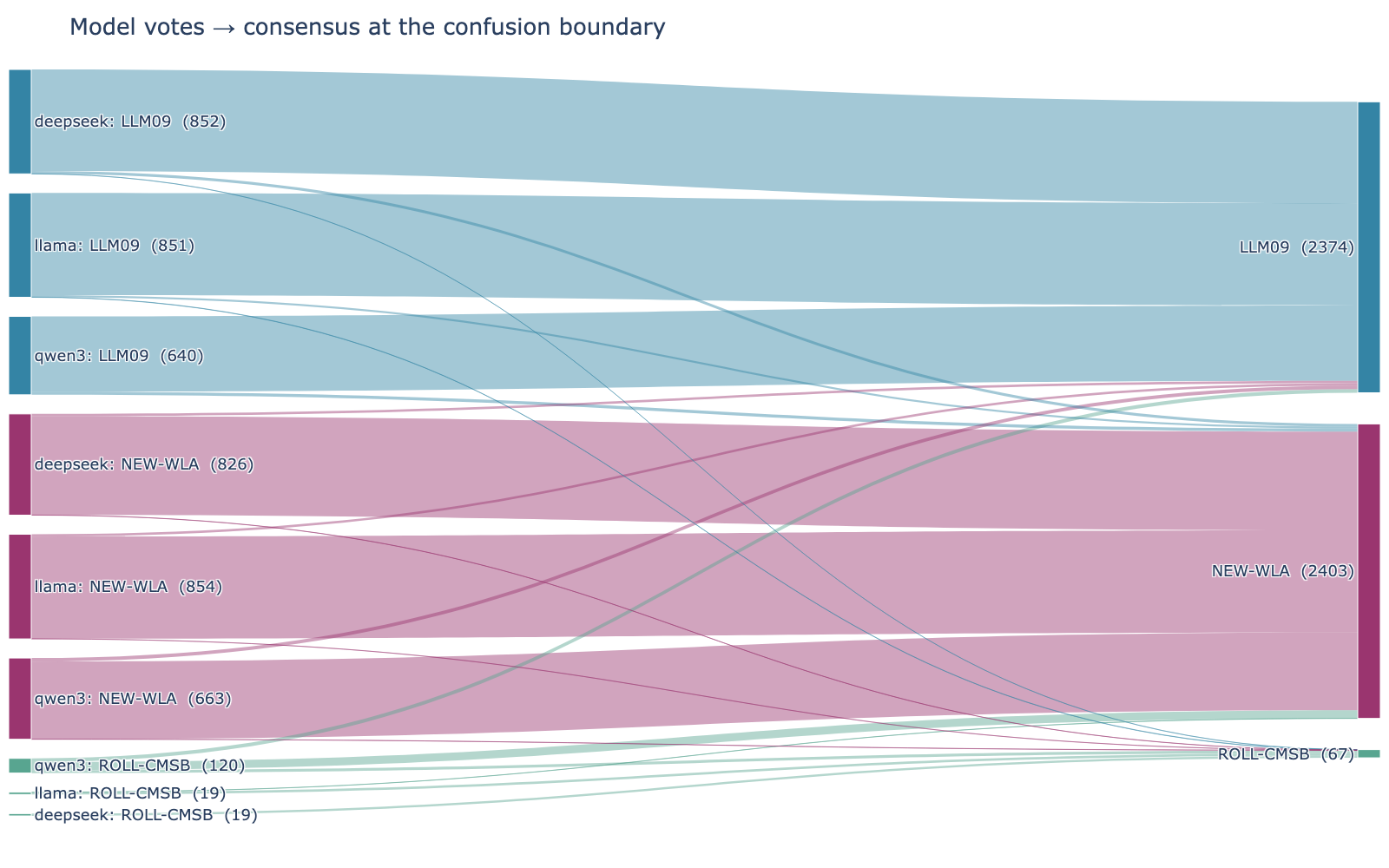}
\caption{Flow of incidents across the LLM09 / NEW-WLA / ROLL-CMSB confusion
boundary.}
\end{figure}

\begin{wrapfigure}{R}{0.45\textwidth}
\centering
\includegraphics[width=0.42\textwidth]{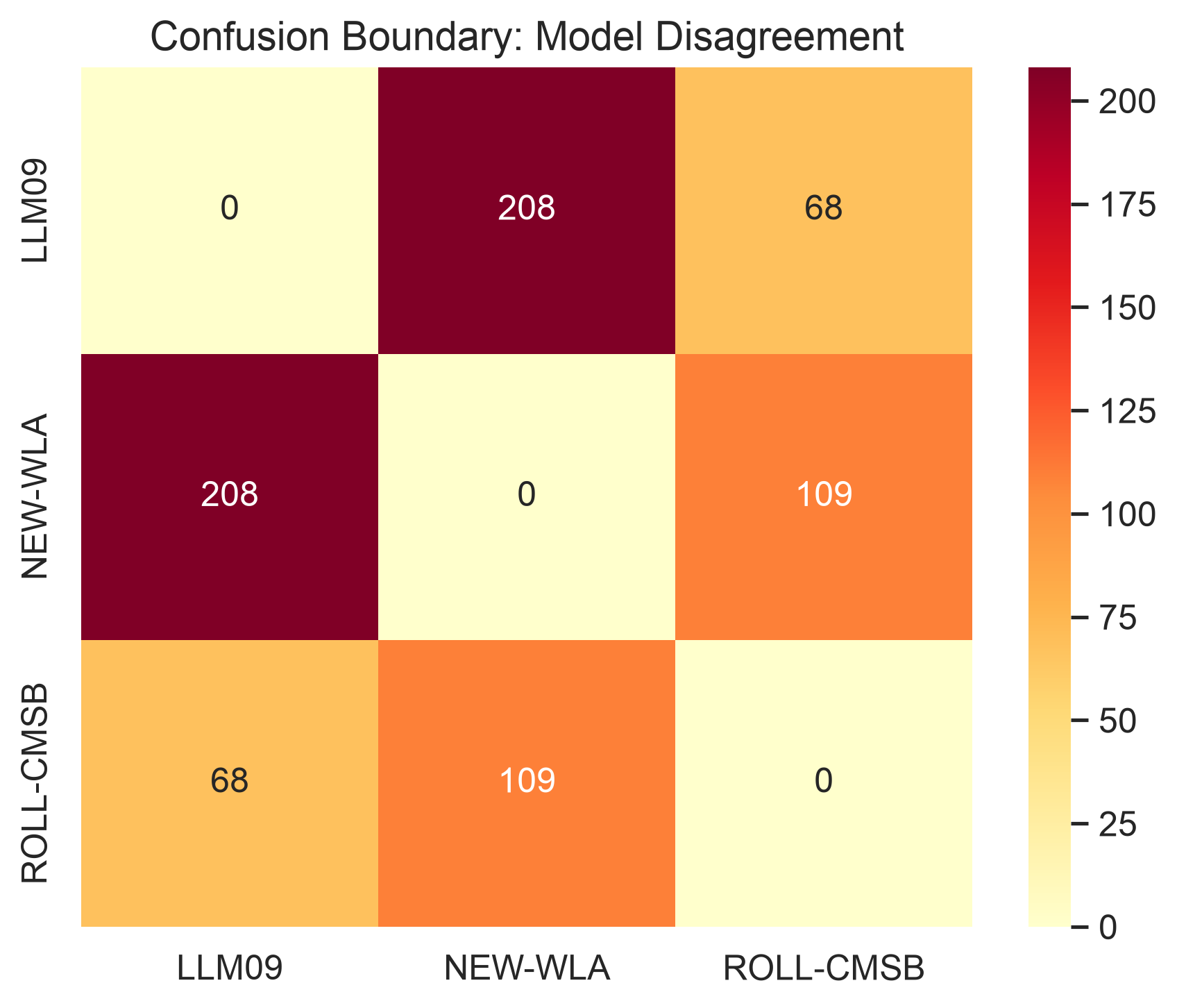}
\caption{Three-by-three confusion among LLM09, NEW-WLA, and ROLL-CMSB.}
\end{wrapfigure}

\subsection{Act 10: What Part II shows}\label{act-10-what-part-ii-shows}

Where the data and the experts agree, confidence is highest. LLM02
(Sensitive Information Disclosure) sits near the top of both. ROLL-SICG,
NEW-ITSCD, and NEW-MSDA sit near the bottom of both. These positions
hold across the uncertainty ranges, and they are the safest to act on.

Where the data pushes back, it does so for readable reasons. LLM09's
incident volume far exceeds its expert rank, driven partly by a broad
definition that sweeps in AI-adjacent harms and inflated by the LLM09 /
NEW-WLA / ROLL-CMSB confusion boundary, which makes all three counts
less reliable than entries with cleaner definitions. NEW-WLA shows the
same pattern.

Where the experts see what the incidents miss, the gap is about timing.
NEW-PMP and NEW-MTIE have strong expert signal and almost no public
incidents, because the threats are new. For emerging risks the expert
signal leads, which is exactly what a quarter-weight on the data is
meant to allow.

The method triangulates two imperfect signals; it does not hand down a
verdict from either. The incident data has structural biases: a sampling
frame that misses unreported incidents, measured classifier error, and
taxonomy-frame circularity, which means we are partly measuring the
classifier's preferences rather than the true threat distribution. The
expert vote has its own biases: availability, recency, anchoring to last
year's list. The value is in the comparison. Where they agree,
confidence rises; where they diverge, the divergence itself is the
finding.

Stated plainly: the incident data agrees with the expert ranking only
weakly (Cohen's κ ≈ 0.20, with an interval that crosses zero), and ---
as Part III shows next --- the expert ranking is robust: four frontier
classifiers and a ground-truth check do not move it. Weak agreement and
a stable ranking are not in tension. A quarter-weight corrective that
agrees weakly is doing exactly what it was designed to do, and a ranking
a stronger classifier cannot improve is one to rely on while the
measurement gets better.

\section{Part III --- Robustness Under Frontier
Classifiers}\label{part-iii-robustness-under-frontier-classifiers}

\subsection{Act 11: Does a better classifier change the
ranking?}\label{act-11-does-a-better-classifier-change-the-ranking}

The incident-derived ranking depends on the classifier that labeled the
corpus. If a stronger model would reorder the list, the ranking is an
artifact of a weak classifier and should not be trusted. So we
pre-registered a bake-off (\citeproc{ref-nosek2018}{Nosek et al. 2018}):
four frontier models re-labeled the evaluation set, and the rule for
declaring a winner was fixed before any model ran. A model had to beat
the 2026 incidence floor on balanced accuracy and clear a significance
test.

\begin{quote}
\textbf{Sidebar --- balanced accuracy.} Plain accuracy flatters a
classifier on lopsided data: a model that always guesses the common
class scores high while missing every rare one. Balanced accuracy
averages the recall within each class, so a rare category counts as much
as a common one. On this kind of task it runs from about 0.5 (chance) to
1.0 (perfect) (\citeproc{ref-brodersen2010}{Brodersen et al. 2010}).
\end{quote}

The bake-off returned no winner. The 2026 floor scored balanced accuracy
0.863. All four frontier models scored below it: llama-405b 0.744
(\citeproc{ref-llama3}{{Grattafiori et al.} 2024}), qwen3-235b 0.733
(\citeproc{ref-qwen3}{{Yang et al.} 2025}), deepseek-v3 0.711
(\citeproc{ref-deepseekv3}{DeepSeek-AI 2024}), mistral-large-2411 0.691
(\citeproc{ref-mistrallarge2411}{Mistral AI 2024}). None cleared the
floor, so under the pre-registration the 2026 ranking stands. (Every
number in this section is read from
\texttt{cycles/2026-rarr/results/robustness\_validation.json} and the
RARR conclusion beside it.)

A null result, no model beating the floor, could mean the floor is
genuinely good or that the test is weak. So we checked the floor against
ground truth directly. Scored against the adjudicated gold set as truth,
the floor's incidence ranking correlates with the true ranking at
Spearman ρ = 0.918.

\begin{quote}
\textbf{Sidebar --- Spearman ρ.} Spearman's rho measures how well two
rankings agree on order, ignoring the exact scores. It runs from −1 (one
ranking is the reverse of the other) through 0 (no relationship) to +1
(identical order). A ρ of 0.918 means the floor's ordering is very close
to the true ordering (\citeproc{ref-spearman1904}{Spearman 1904}).
\end{quote}

Every frontier model's apparent edge over the floor dissolves under a
bootstrap.

\begin{quote}
\textbf{Sidebar --- bootstrap.} The bootstrap estimates how much a
number would wobble if the data had come out slightly differently. It
re-samples the observations with replacement many times, recomputes the
number on each re-sample, and reads the spread as a confidence interval.
When that interval crosses zero, the apparent effect is within noise
(\citeproc{ref-efron1979}{Efron 1979}).
\end{quote}

The paired-bootstrap interval for each model's ranking-fidelity gain
over the floor crosses zero: llama-405b +0.001 (95\% interval −0.047 to
+0.055), the ensemble +0.009 (−0.032 to +0.050), mistral-large-2411
−0.012 (−0.053 to +0.028). Reweighting the gold set to the corpus's
class mix raises the floor to ρ = 0.971 and leaves the best frontier
configuration statistically tied (interval −0.025 to +0.024). The figure
shows each classifier's ranking fidelity against the floor.

\begin{wrapfigure}{L}{0.51\textwidth}
\centering
\includegraphics[width=0.48\textwidth]{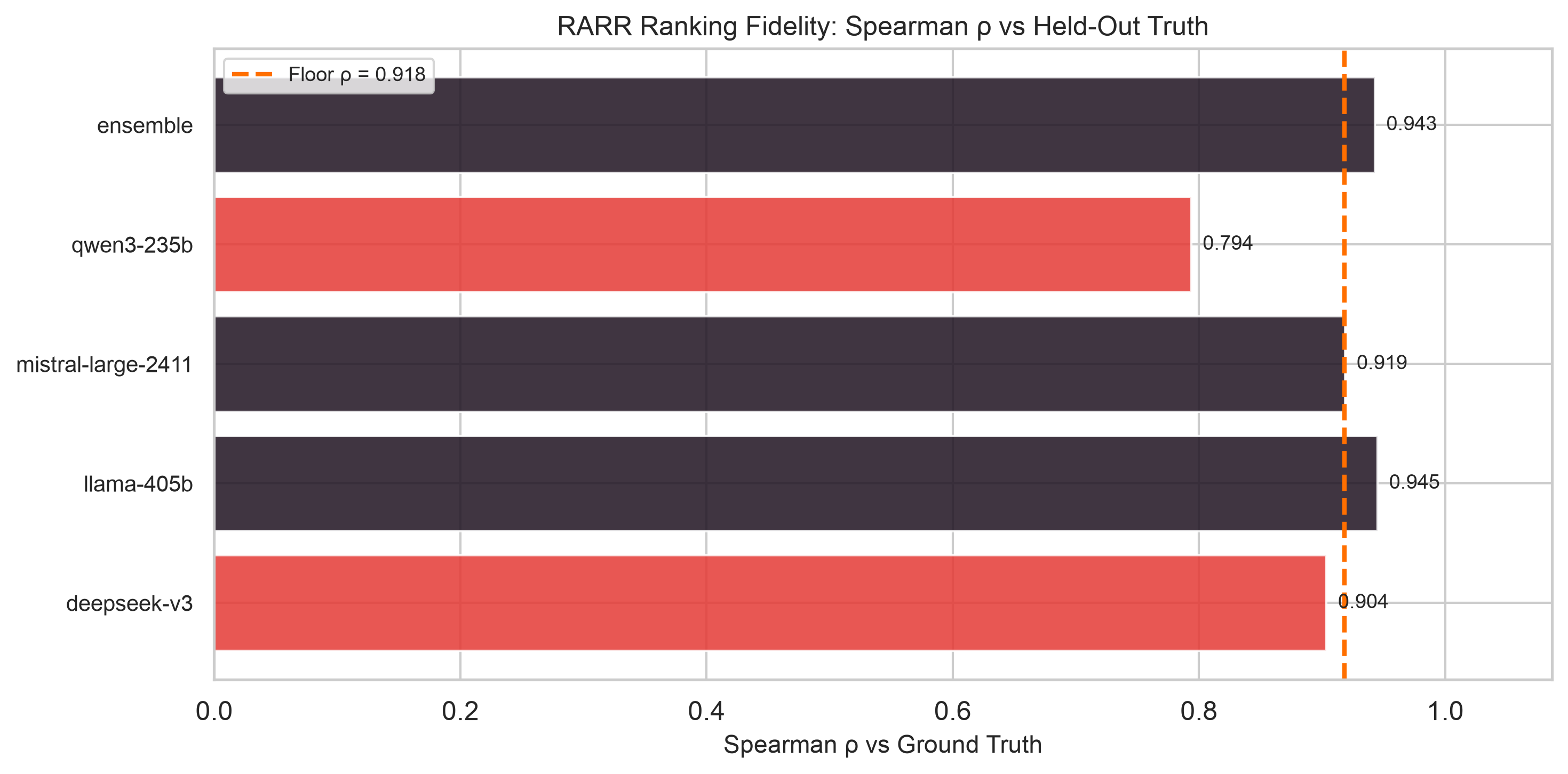}
\caption{Ranking fidelity (Spearman ρ against held-out truth) for each frontier
classifier and the ensemble, against the 2026 incidence floor. Every
frontier gain over the floor has a bootstrap interval that crosses zero.}
\end{wrapfigure}

\subsubsection{The recall correction, and its one honest
limit}\label{the-recall-correction-and-its-one-honest-limit}

One difference between the floor and the frontier models survives
correction. The floor classifier over-files incidents into a few crowded
categories (LLM02, LLM09, ROLL-CMSB), which inflates their raw counts.
The pipeline's recall-and-precision correction is built to remove
exactly this kind of distributional gap, and it does: after correction,
the floor and the ensemble reach effectively the same per-class
magnitude accuracy (cross-validation-corrected neg-L2 −0.0024 for the
floor, −0.0033 for the ensemble; the ensemble-minus-floor difference has
a 95\% interval of −0.0009 to +0.0007, which crosses zero).

The correction has a limit it cannot cross. The 2026 classifier never
predicts ``out of scope'': it assigns a specific taxonomy category to
every incident, including the roughly 38\% of the gold set that is truly
out of scope. Its out-of-scope recall is therefore 0\%, against 0.49 for
a frontier model on the same set. A recall correction adjusts for how
often a classifier misses a class it does predict; it cannot recover a
class the classifier never predicts at all. On this one axis the
frontier models are genuinely better: they spot out-of-scope incidents
that the floor files into a category. This changes the magnitudes of a
few per-class counts. It does not change the rank order: the reweighting
and bootstrap tests above already price in the false-positive inflation,
and the ordering holds.

The conclusion of Part III is narrow and firm. No frontier classifier
reorders the 2026 ranking, on either the ordinal order or the
recall-corrected magnitudes. The ranking is robust to classifier choice.
The one measured advantage of the frontier models, out-of-scope
detection, changes magnitudes the pipeline already corrects for, not the
order of the list.

\subsection{The 2026 blended Top 10}\label{the-2026-blended-top-10}

The blend combines two witnesses into a distribution over each risk's
final position. The expert vote carries three-quarters weight, the
incident data one quarter. Each risk gets a spread of plausible
positions, and the ten fall into three tiers.

\begin{quote}
\textbf{Sidebar --- distribution over positions.} Each posterior draw
pairs one sample of the incident rates with one sample of the expert
ranking, blends them, and reads off a position. Sixteen thousand draws
give a distribution over positions for each risk, so a firm placement
and a coin flip look different on the page.
\end{quote}

\begin{wrapfigure}{L}{0.73\textwidth}
\centering
\includegraphics[width=0.70\textwidth]{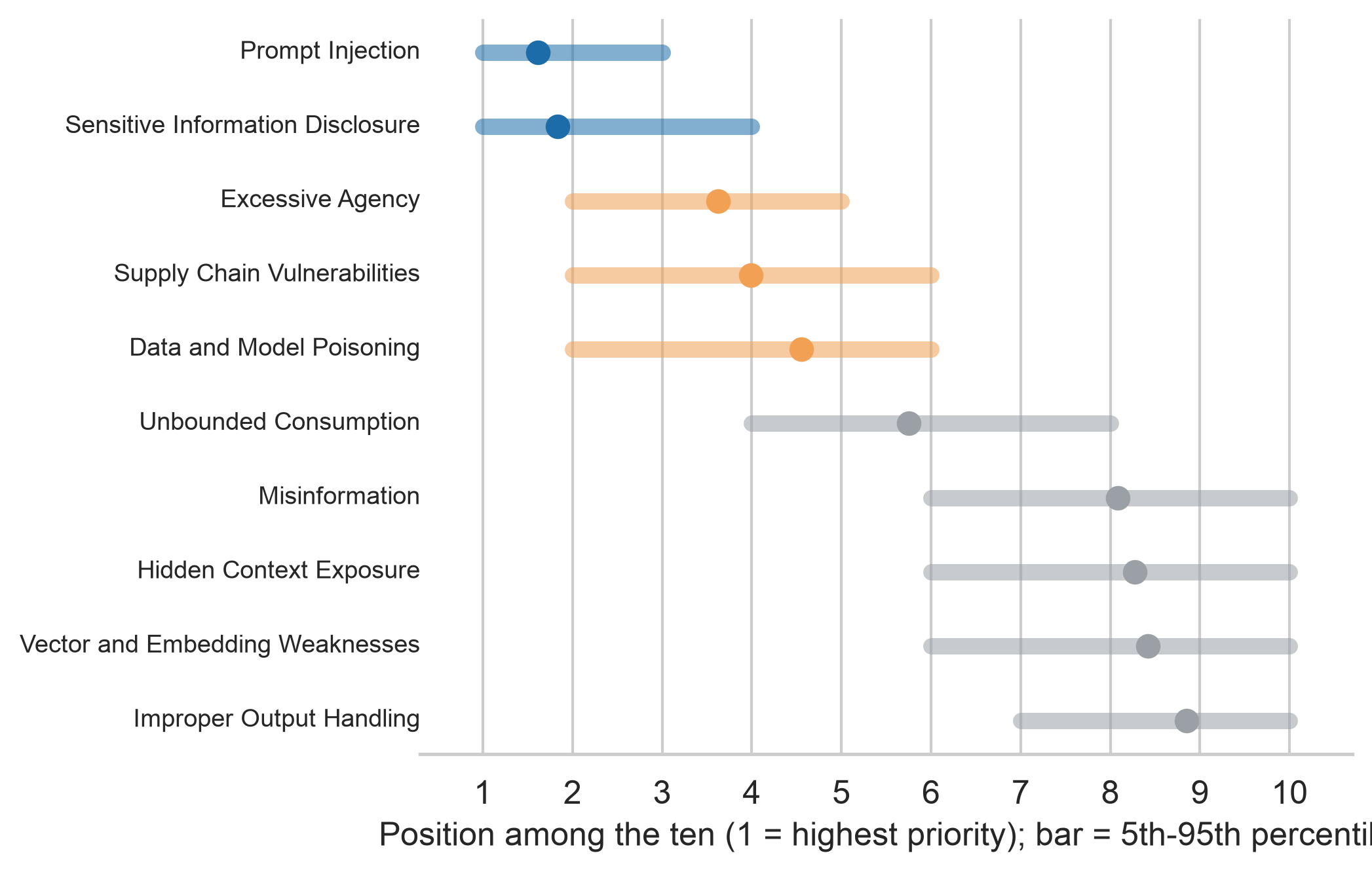}
\caption{Each risk at its mean position, with a bar spanning the 5th to 95th
percentile. Three tiers read directly: a tight pair, an overlapping
band, and a wide tail.}
\end{wrapfigure}

\textbf{The co-leading pair.} Prompt Injection and Sensitive Information
Disclosure hold the top, each a near-certain top-three risk (P(top-3)
0.99 and 0.95). The two blends disagree on which ranks first: the
simpler rank-space blend puts Sensitive Information Disclosure ahead,
the probabilistic blend puts Prompt Injection ahead, and their intervals
overlap. We report them as co-leading, with no method-independent first
place.

\textbf{The tied band.} Excessive Agency, Supply Chain, and Data and
Model Poisoning form a middle band with overlapping intervals. Excessive
Agency moves up into the band from its published 2025 position, the one
clear mover here.

\textbf{The tail.} Unbounded Consumption sits at the top of the tail,
placed by the expert vote alone (its incident recall the corpus cannot
estimate), and it reaches the top five in about a third of the draws.
The remaining four --- Misinformation, Hidden Context Exposure, Vector
and Embedding Weaknesses, Improper Output Handling --- reach the top
five in under one draw in twenty. We present the tail as a group and do
not report an order inside it.

\begin{quote}
\textbf{Sidebar --- why a distribution beats a single rank number.} A
single rank hides its own uncertainty. Two adjacent tail positions
differ by a fraction of a place across the draws, so a printed rank
would claim precision the data does not carry. The tiers report only
what the spread supports.
\end{quote}

\begin{figure}[htbp]
\centering
\includegraphics[width=0.62\textwidth]{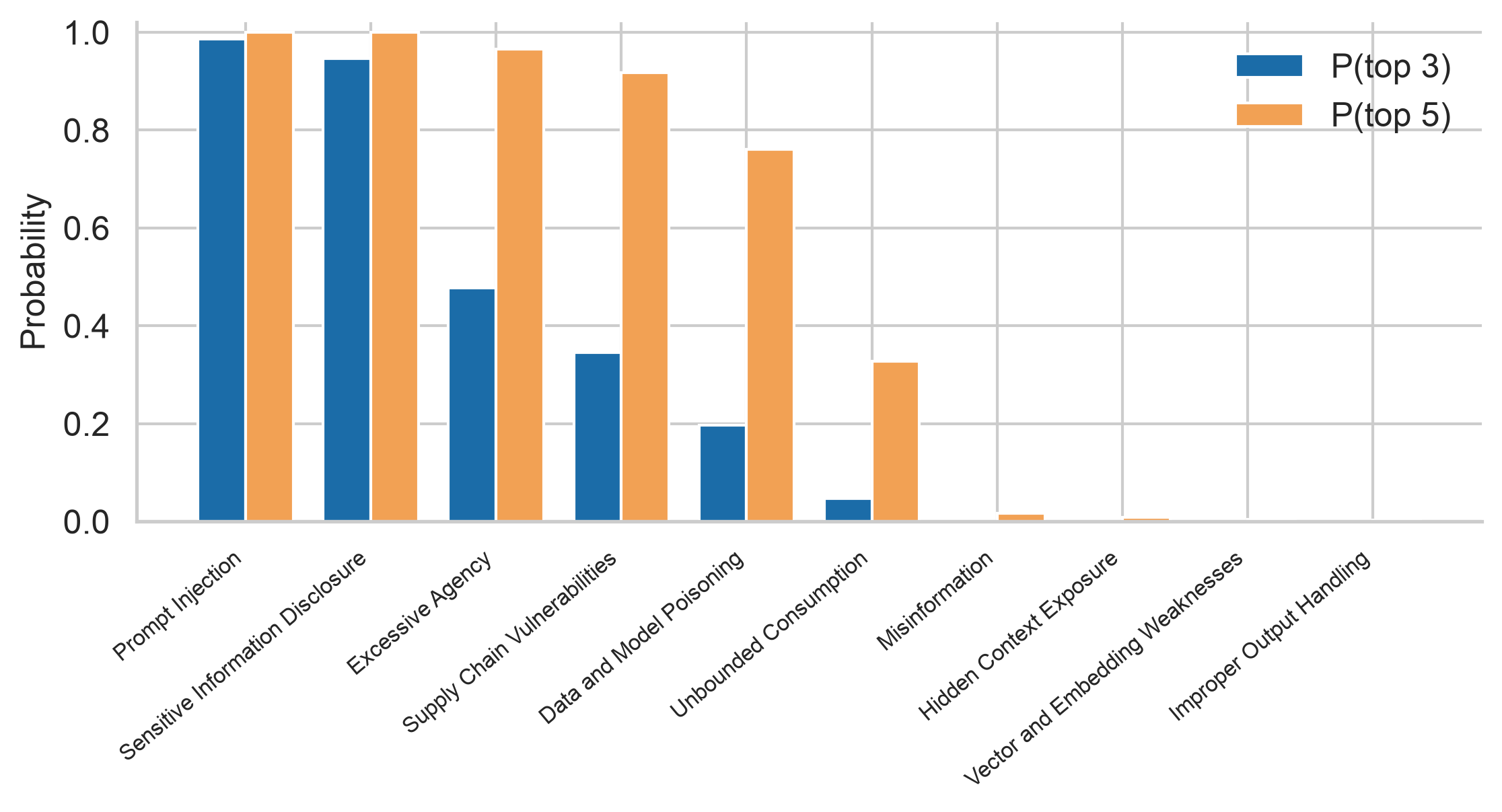}
\caption{The probability each risk reaches the top three and the top five, per
tier. The co-leading pair sits near certainty; the tail sits near zero.}
\end{figure}

A simpler rank-space blend, which uses only the order of each witness
and discards the magnitudes, gives nearly the same bulk ordering
(Kendall's tau (\citeproc{ref-kendall1938}{Kendall 1938}) computed at
run time, about 0.87). The one place the two methods disagree is the
top, which is why we present the top two as a pair.

\textbf{Against the published list.} The OWASP GenAI LLM Top 10 2026
published in August 2026, after this analysis was run
(\citeproc{ref-owasp2026llmtop10}{OWASP GenAI Security Project 2026}).
Its ten entries follow the tiers reported here: Prompt Injection and
Sensitive Information Disclosure at the top, then Excessive Agency,
Supply Chain, and Data and Model Poisoning, then Unbounded Consumption
ahead of Misinformation, Hidden Context Exposure, Vector and Embedding
Weaknesses, and Improper Output Handling. The published list assigns a
position inside the tail. This report does not, because the draws do not
separate those four risks. That difference is one of presentation, and
it does not touch the tiers above it.

\subsection{Glossary}\label{glossary}

Every term defined in a sidebar, collected here for reference.

\textbf{0.75 / 0.25 blend.} The rule that combines the expert vote and
the incident data into one ranking. Each risk's expert rank and incident
rate are placed on a common scale (a z-score) and combined at fixed
weights, three-quarters vote and one quarter data, once per posterior
draw. Sorting the resulting scores highest-first for each draw produces
a distribution over positions for every risk, summarized here as three
tiers.

\textbf{Balanced accuracy.} Accuracy that averages the recall within
each class, so a rare category counts as much as a common one. It avoids
the trap of plain accuracy, which a model can inflate by always guessing
the common class. On this task it runs from about 0.5 (chance) to 1.0
(perfect).

\textbf{Blind labeling.} Recording a human judgment before seeing the
machine's answer, so the human is not anchored to it. The gold-set
reviewer labeled each incident from its text alone, then revealed the
model votes, then decided.

\textbf{Bootstrap.} A way to gauge how much a number would wobble under
slightly different data. Re-sample the observations with replacement
many times, recompute the number each time, and read the spread as a
confidence interval. An interval that crosses zero means the apparent
effect is within noise.

\textbf{Classifier.} Any procedure that reads an input and assigns it to
one of a fixed set of categories. Ours is an ensemble of three language
models voting on which taxonomy entry (or ``out of scope'') each
incident belongs to. It is a measuring instrument, not ground truth.

\textbf{Cohen's κ (kappa).} Agreement between two labelings after
subtracting the agreement expected by chance. κ = 1 is perfect, κ = 0 is
chance-level, negative κ is systematic disagreement. When its
uncertainty interval crosses zero, the agreement is weak whatever the
point estimate.

\textbf{Credible interval.} The Bayesian range of plausible values for a
quantity. A 90\% credible interval holds 90\% of the posterior's
probability, so the true value sits inside it with 90\% probability
given the model and data. Wide means uncertain.

\textbf{Credible interval over a rank.} A credible interval computed on
an entry's rank position, the same idea used elsewhere in this report
for a count-scale quantity like λ. It gives the middle span of plausible
ranks a risk could hold under the model and the data. The interval
assumes a single correct rank is the value it brackets, an assumption
that holds less firmly for the three frame-blind entries, whose rank
comes from the vote alone.

\textbf{Distribution over positions.} The spread of positions a risk can
land at across many posterior draws. Each draw pairs one incident-rate
sample with one expert-rank sample, blends them, and places the risk at
a position. Collecting the position from all 16,000 draws produces a
full spread for each risk, from a tight cluster near one place to a wide
scatter across many.

\textbf{Frame-blind drop.} The adjustment applied to the three
frame-blind entries (Data and Model Poisoning, Vector and Embedding
Weaknesses, Unbounded Consumption): their weight shifts from
three-quarters vote and one quarter data to full vote weight, so the
blend places them from the expert rank alone. Their incident rates still
enter the shared scale used to score every other risk. The shift changes
where a frame-blind entry lands, and the report states that plainly.

\textbf{Gold set.} A batch of records labeled carefully by a human to
serve as the answer key --- here, 1,200 adjudicated incidents used both
to calibrate the classifier and as the ground truth for the robustness
check.

\textbf{Incident corpus.} A fixed, documented collection of incident
records assembled for analysis. Ours is a dated snapshot of publicly
reported LLM-security incidents, each a short text description, drawn
from public databases.

\textbf{Kendall's tau.} A second measure of how well two rankings agree
on order, alongside Spearman's ρ. It checks every pair of entries and
counts how often the two rankings place them in the same relative order,
turning that count into a score from −1 (every pair reversed) through 0
(no relationship) to +1 (every pair agrees). This report uses it to
compare the probabilistic blend's bulk order against a simpler
rank-space blend that keeps only the ranks and discards the underlying
scores.

\textbf{Latent incidence (λ).} The true, unobserved underlying rate at
which incidents of a category occur, as distinct from the raw count the
classifier reported. The model infers λ from the noisy counts after
correcting for precision and recall.

\textbf{MCMC (Markov chain Monte Carlo).} A method for exploring a
probability distribution too complex to write down in closed form, by a
guided random walk that spends more time where the answer is more
plausible. The collected steps approximate the posterior; we drew
16,000.

\textbf{Negative-binomial measurement-error model.} A count model with
two features: it treats classifier counts as noisy measurements of the
true rate and corrects for the noise (measurement-error), and it uses
the negative-binomial distribution, which --- unlike the Poisson ---
lets the spread of counts exceed their average (over-dispersion).

\textbf{Out-of-scope.} The category for incidents that belong to none of
the 20 taxonomy entries: a real AI harm that is not a vulnerability in a
large language model. Marking an incident out of scope is a correct
classification.

\textbf{P(top-k).} Across the 16,000 posterior draws, the share that
place a risk within the top k positions (k = 3 or 5 in this report). A
risk that is near-certain to rank highly shows a P(top-3) close to 1.
The three-quarters vote weight dominates this probability wherever the
vote and the data point the same direction, so a high P(top-k) usually
traces back to a strong expert-vote placement that the incident data
does not contradict.

\textbf{Precision.} Of the incidents the classifier files under a
category, the fraction that truly belong there. Low precision means most
of what lands in a category does not belong to it.

\textbf{Prior and posterior.} The prior is what the model assumes before
seeing this data; the posterior is the updated belief after combining
prior and data. The posterior is a distribution of plausible values, not
a single number.

\textbf{Probabilistic blend.} The method this report uses for the 2026
ranking: combine the incident-rate posterior and the expert-vote
posterior in score space, once per posterior draw, and read off a
position for each risk. Repeating this over 16,000 draws produces a
distribution over positions for each risk. The report summarizes that
distribution as a mean position, a P(top-k), a credible interval over
each rank, and three tiers.

\textbf{Recall.} Of the incidents that truly belong to a category, the
fraction the classifier finds. A classifier can have high precision and
low recall, or the reverse; the two are measured separately.

\textbf{Spearman ρ (rho).} How well two rankings agree on order,
ignoring exact scores. It runs from −1 (reversed) through 0 (unrelated)
to +1 (identical order).

\textbf{Sum-prevalence fold.} The rule for combining a rolled-up child
entry's incident rate into its parent on the data axis: the child's rate
adds to the parent's, so incidents recorded against the child still
count toward the parent's total. Four entries fold this way in the 2026
blend. The vote axis folds by a separate rule: the parent keeps the
better of its own rank and its child's rank, since votes do not add the
way rates do.

\textbf{Tied tier.} A group of entries whose plausible-position
intervals overlap enough to fold into one block for reporting purposes.
The 2026 blend sorts the ten incumbents into three such tiers: a
co-leading pair at the top, a tied band in the middle, and a wide tail
with no defended order inside it. Membership reflects overlapping
position intervals: the draws place these risks so close together that
the ranking does not separate them.

\subsection{Limitations and Independent-Review
Status}\label{limitations-and-independent-review-status}

This is an internally rigorous analysis with real limits. We state them
plainly rather than bury them.

\textbf{The gold set is single-author.} One reviewer (the project
author) adjudicated all 1,200 gold-set incidents that calibrate the
classifier and serve as the ground truth in Part III. The reviewer's
independent blind label disagreed with the model consensus at a rate of
0.75, and the final adjudication overrode the consensus on 553 of the
1,200 incidents. Those rates are high, and they reflect genuinely
ambiguous categories rather than a broken pipeline. A single annotator
cannot measure inter-rater reliability (\citeproc{ref-cohen1960}{Cohen
1960}); a second independent adjudicator would harden the truth target.
The ranking-fidelity bootstrap median of exactly 0.000 makes a hidden
improvement unlikely regardless, but the single-author gold set remains
the central limitation.

\textbf{The reviewers are interim.} The rubric reviewer and the
statistical reviewer for this analysis are, at this stage, the ranking
author. Independent adjudication of the gold set and independent
statistical review are future work, not completed steps. Read the
findings as exploratory and pending independent adjudication.

\textbf{The agreement signal is weak and uncertain.} Cohen's κ between
the vote and the incident data is 0.20 with a 90\% interval of −0.16 to
0.57. The interval crosses zero, so we cannot rule out that the two
rankings agree only by chance. The honest bottom line: weak agreement,
not confirmation.

\textbf{The corpus has known blind spots.}

\begin{itemize}
\tightlist
\item
  \emph{Stratum imbalance.} The labeled corpus is 6,297 security
  incidents against 342 ai-harm incidents. Precision was hand-verified
  only on the security stratum; ai-harm precision falls back to an
  uninformative prior. Conclusions about ai-harm categories rest on
  weaker measurement.
\item
  \emph{The out-of-scope blind spot.} The base classifier never predicts
  ``out of scope'' and files every incident into some category,
  including the roughly 38\% of the gold set that belongs in none. Part
  III shows the ranking survives the resulting false-positive inflation,
  but the raw per-class counts overstate the crowded categories.
\item
  \emph{Frame-blind entries.} Three entries --- LLM04, LLM08, LLM10 ---
  draw their signal from a single stratum, so the corpus cannot
  cross-check their recall, and their data ranks are soft by
  construction. Several broad categories (LLM09, NEW-WLA, ROLL-CMSB)
  also sit on confusion boundaries the classifier cannot cleanly
  resolve, so their counts are less reliable than entries with sharp
  definitions.
\end{itemize}

\textbf{Status.} This report is exploratory and internally rigorous. It
is not peer-reviewed, and it is not the official OWASP release. Full
external publication would require an independent adjudicated gold set,
independent statistical review, and a broader corpus that narrows the κ
interval. Until then the blended ranking is a working reconciliation of
two imperfect signals, offered for scrutiny.

\subsection{Data and Code
Availability}\label{data-and-code-availability}

The engine, the analysis notebook, and the artifacts behind every number
in this report are public at
https://github.com/rocklambros/incident-rank-validation . The repository
licenses in three parts: the software under Apache-2.0, this report and
the notebooks under CC BY-SA 4.0, and the data artifacts under CC BY
4.0. Every file falls into exactly one part, and the repository NOTICE
is the authoritative statement.

\textbf{What is published.} The labeled corpus holds 6,639 records in
\texttt{projects/owasp-llm/cycles/2026/classify/labeled\_incidents.json},
each carrying its taxonomy entry, the classifier's confidence and
rationale, and its stratum. Incident text for 6,142 of those records is
committed in \texttt{calibration/llm\_prelabels.jsonl} alongside the
three-model votes. The gold set is published in full at
\texttt{calibration/adjudicated\_goldset.jsonl}: all 1,200 incidents,
with the reviewer's blind label, the model consensus, and the final
adjudication as separate fields, so the disagreement rates reported
above can be recomputed rather than taken on trust. Per-incident
predictions from the four frontier classifiers in Part III are under
\texttt{cycles/2026-rarr/classify/seq/}.

\textbf{What reproduces.} The pre-registration manifest
(\texttt{cycles/2026/prereg/manifest.json}) fixes the model
specification, the taxonomy hash, the corpus snapshot hash, and the PRNG
seed, all recorded before any result was read. Posteriors, precision
verification, and the blended ranking are committed as computed
artifacts, so every figure and statistic regenerates from the repository
without re-running the classifier against paid APIs. Build instructions
are in \texttt{notebooks/preprint/BUILD.md}.

\textbf{What is not published.} Incident records carry internal
identifiers of the form \texttt{INC-04490}, not upstream CVE, GHSA, OSV,
or AIAAIC keys. A reader cannot join these labels back to the source
databases record by record. Text is absent for 497 of the labeled
records. The source databases retain their own terms and remain the
authority for the underlying incident descriptions; what we redistribute
is derived labels and snapshots.

\subsection{Scope and Authority}\label{scope-and-authority}

This report is an incident-data analysis authored by two members of the
OWASP working group. It is not the official OWASP Top 10 for LLM
Applications, it does not supersede the official list or the process
that produces it, and it does not speak for OWASP.

The OWASP GenAI LLM Top 10 2026 published in August 2026
(\citeproc{ref-owasp2026llmtop10}{OWASP GenAI Security Project 2026}).
This analysis ran during that cycle, against the working group's ranking
as it stood before publication, and ``the 2026 list'' throughout this
report means that ranking. This report stress-tests it against incident
data. It does not set it. Where the two coincide, the published document
is the authority: it is the official artifact, and this report is a
robustness check on the reasoning behind it, not a substitute for it.

The work is exploratory and internally rigorous, not a peer-reviewed
finding. Its value is transparency and a robustness check: it shows how
a community-expert ranking holds up when confronted with a large-scale
incident corpus, and it is candid about where the data is weak. Read it
as a stress test offered for scrutiny, not as an authority on the
ranking.

\section*{References}\label{bibliography}
\addcontentsline{toc}{section}{References}

\protect\phantomsection\label{refs}

\end{document}